\documentclass[11pt]{article}

\usepackage{times,epsfig,epsf,psfrag,amsmath,amssymb,graphicx,graphics,verbatim}
\usepackage{makeidx}
\usepackage{moreverb}
\usepackage[ruled,vlined]{algorithm2e}
\usepackage[usenames]{color}

\newcommand{\LL}{\textsc{level}}
\newcommand{\OO}{{\cal O}}
\newcommand{\MM}{{\cal M}}
\newcommand{\EE}{{\cal E}}
\newcommand{\EX}{{\mathbf{E}}}

\newcommand{\PP}{\mathbf{Pr}}

\newcommand{\ignore}[1]{{}}

\newenvironment{proof}{\par\noindent{\bf Proof:}}{\mbox{}\hfill$\Box$\\}

\newtheorem{definition}{Definition}[section]

\newtheorem{theorem}{Theorem}[section]

\newtheorem{lemma}{Lemma}[section]
\newtheorem{remark}{Remark}[section]

\newtheorem{observation}{Observation}[section]
\newtheorem{corollary}{Corollary}[theorem]

\SetKwComment{tcc}{/*}{*/}
\SetKwComment{tcp}{//}{}
\SetKwComment{Comment}{}{}
\SetKwIF{If}{ElseIf}{Else}{if}{then}{else if}{else}{endif}
\SetKwFor{While}{while}{do}{endw}
\SetKwFor{For}{for}{do}{endfor}
\SetKwFor{ForEach}{foreach}{do}{endfch}

\begin{document}

\title{Fully dynamic maximal matching in $O(\log n)$ update time 
%:\\ with revised and corrected analysis
%\footnote{A preliminary version of this work
%appeared in IEEE FOCS 2011}
}

\author{Surender Baswana\\
   Department of CSE,\\
   I.I.T. Kanpur, India\\
   {\small\texttt{sbaswana@cse.iitk.ac.in}}
 \and
Manoj Gupta\\
   Department of CSE,\\
   I.I.T. Delhi, India\\
{\small\texttt{gmanoj@cse.iitd.ernet.in}}
\and
Sandeep Sen\\
   Department of CSE,\\
   I.I.T. Delhi, India\\
  {\small\texttt{ssen@cse.iitd.ernet.in}}
}
\maketitle
\begin{abstract}
We present an algorithm for maintaining a maximal matching in a graph under
addition and deletion of edges. Our algorithm is randomized and it
takes expected amortized $O(\log n)$ time for each edge update where $n$ is
the number of vertices in the graph. Moreover, for any sequence of $t$ edge updates, the
total time taken by the algorithm is $O(t\log n + n \log^2 n)$ with high probability.
%
%While there exists a trivial $O(n)$ time
%algorithm for each edge update, the previous best known result for this
%problem is due to Ivkovi\'c and Llyod\cite{IvkovicLlyod94}. For a graph with
%$n$ vertices and $m$ edges, they gave an $O( {(n+ m)}^{0.7072})$ update time
%algorithm which is sublinear only for a sparse graph.
%
%For the related problem of maximum matching, Onak and Rubinfeld
%\cite{OnakRubinfeld10} designed a randomized algorithm that achieves
% expected amortized $O(\log^2 n)$ time for each update for maintaining a
%$c$-approximate maximum matching for some unspecified large constant $c$.
%In contrast, we can maintain a factor two approximate maximum matching in
% expected amortized $O(\log n )$ time per update as a direct corollary of the
%maximal matching scheme. This in turn also implies a
%two approximate vertex cover maintenance scheme that takes
%expected amortized $O(\log n )$ time per update.\\
\end{abstract}

\vspace{1cm}
\noindent
{\bf Note:}~The previous version of this result appeared in SIAM J. Comp., 44(1): 88-113, 2015. However, the analysis
presented there for the algorithm was erroneous. This version rectifies this deficiency 
without any changes in the algorithm while preserving the performance bounds of the original algorithm.

\pagebreak
\section{ Introduction }
Let $G=(V,E)$ be an undirected graph on $n=|V|$ vertices and $m=|E|$ edges.
A matching in $G$ is a set of edges $M\subseteq E$ such that no two 
edges in $M$ share any vertex. The study of matchings satisfying various 
properties has remained the focus of graph theory for decades \cite{LP86}. It is
due to the elegant structure of matching that it also appears in 
various combinatorial optimization problems \cite{EdmondsJohnson73,Lawler76}. 
A few well studied matching problems are maximum cardinality matching 
\cite{Edmonds65,HopcroftKarp73,MicaliVazirani80,MuchaS04}, maximum weight 
matching \cite{GabowTarjan91,HuangKavitha12}, minimum cost matching 
(chapter 7, \cite{KleinbergTardos05}), stable matching \cite{GaleShapley62}, 
popular matching \cite{AbrahamIKM07}. Among these problems, the maximum 
cardinality matching problem has been studied most extensively.
A matching $M$ is of maximum cardinality if 
the number of edges in $M$ is maximum. A maximum cardinality matching
(MCM)
is also referred to as a {\it maximum matching}. A matching is said to be a 
{\it maximal} 
matching if it cannot be strictly contained in any other matching. It is well 
known that a maximal matching guarantees a 2-approximation of the maximum 
matching. Though it is quite easy to compute a maximal matching in $O(m+n)$ 
time, designing an efficient algorithm for maximum matching has remained a 
very challenging problem for researchers \cite{Edmonds65,MicaliVazirani80}. 
The fastest algorithm, till date, for maximum matching runs in $O(m\sqrt{n})$ 
time and is due to Micali and Vazirani \cite{MicaliVazirani80}. In this 
paper, we address the problem of maintaining a maximal matching in a dynamic graph.

Most of the graph applications in real life deal with graphs that are not 
static, viz., the graph changes over time caused by deletion 
and insertion of edges. This has 
motivated researchers to design efficient algorithms for various graph problems
in dynamic environment. An algorithmic graph problem is modeled in
the dynamic environment as follows. There is an online 
sequence of insertion and deletion of edges and the goal is to update
the solution of the graph problem after each edge update. A trivial way to 
achieve this is to run the best static algorithm for this problem after each
edge update; clearly this approach is wasteful.  
The aim of a dynamic graph algorithm is to maintain some clever data structure 
for the underlying problem such that the time taken to   
update the solution is much smaller than that of the best static algorithm.
There exist many efficient dynamic algorithms for various
fundamental problems in graphs \cite{BaswanaKS12,HolmLT01,RodittyZwick08,RodittyZwick12,Thorup07}. 

Baswana, Gupta and Sen \cite{DBLP:journals/siamcomp/BaswanaGS15} had presented a fully dynamic algorithm
for maximal matching which achieved $O(\log n)$ expected amortized time per edge insertion or deletion.
Moreover, for any sequence of $t$ edge updates, the
total time taken by the algorithm is $O(t\log n + n \log^2 n)$ with high probability.
Their algorithm improved an earlier result of Onak and Rubinfeld \cite{OnakRubinfeld10} who presented a randomized
algorithm for maintaining a $c$-approximate (for some unspecified large constant $c$)
matching in a dynamic graph with $O(\log^2 n)$ expected amortized time for each edge update.

This algorithm also implied a similar result for maintaining a two 
approximate vertex cover. It is also used as a basis 
for maintaining approximate maximum weight matching in a dynamic 
graph \cite{AnandBGS12}.

Unfortunately, the analysis given in \cite{DBLP:journals/siamcomp/BaswanaGS15} has a crucial flaw,
viz., the statement of Lemma 4.10 \cite{DBLP:journals/siamcomp/BaswanaGS15} is not true in general.
Although the bound on the update time was critically dependent on  
Lemma 4.10, we have presented an alternate proof of the claimed update
time based on an interesting property of the original algorithm that
was not reported earlier.
For completeness, we have presented the full details of the algorithm 
of \cite{DBLP:journals/siamcomp/BaswanaGS15} and the new corrected analysis in this paper. 
This property is interesting in its own right and may have other useful
applications of the algorithm.

\section{An overview}
\label{sec:prob}
Let $M$ denote a 
matching of the given graph at any moment. Every edge of $M$ is called 
a {\em matched} edge and an edge in $E\backslash M$ is called an 
{\em unmatched} edge. For an edge $(u,v)\in M$, we define $u$ to be 
the {\em mate} of $v$ and vice versa. For a vertex $x$, 
if there is an edge incident to it in the matching $M$, then $x$ is a 
{\em matched} vertex; otherwise it is {\em free} or {\em unmatched}.

In order to maintain a maximal matching, it suffices to ensure 
that there is no edge $(u,v)$ in the graph such that both $u$ and $v$ are 
free with respect to the matching $M$. From this observation,  
an obvious approach will be to maintain the information for each vertex 
whether it is matched or free at any stage. When an edge $(u,v)$ is inserted, 
add $(u,v)$ to the matching if $u$ and $v$ are free. For a case when an 
unmatched edge $(u,v)$ is deleted, no action is required. Otherwise, for both 
$u$ and $v$ we search their 
neighborhoods for any free vertex and update the matching accordingly.  
It follows that each update takes $O(1)$ computation time except when 
it involves deletion of a matched edge; in this case the computation time is of 
the order of the sum of the degrees of the two vertices. 
So this trivial algorithm is quite efficient for {\em small} degree vertices, 
but could be expensive for {\em large} degree vertices. 
An alternate approach to handling deletion of a matched edge is to use a
simple randomized technique - a vertex $u$ is matched with a randomly chosen 
neighbor $v$. Following the standard adversarial model, it can be observed 
that an expected $\deg(u)/2$ edges incident to $u$ will be deleted 
before deleting the matched edge $(u,v)$. 
So the expected amortized cost per edge deletion for $u$ is roughly 
$O\Big( \frac{\deg(u)+\deg(v)}{\deg(u)/2}\Big)$. If $\deg(v)<\deg(u)$, 
this cost is $O(1)$; but if $\deg(v)\gg \deg(u)$, 
then it can be as bad as the trivial 
algorithm. We combine the idea of choosing a random mate and 
the trivial algorithm suitably as follows. 
We first present a fully dynamic algorithm which achieves 
expected amortized $O(\sqrt{n})$ time per update.
We introduce a notion of 
{\em ownership} of edges in which we assign an edge to that endpoint which has 
{\em higher} degree. 
We maintain a partition of the set of vertices into two levels : 0 and 1. 
Level 0 consists of vertices which own  {\em fewer} edges than an 
appropriate threshold and we handle the 
updates in level 0 using the trivial algorithm. Level 1 consists of 
vertices (and their mates) which own {\em larger} number of edges and we use 
the idea of random mate to handle their updates. In particular, a vertex 
chooses a random mate from its set of owned edges which ensures that it selects 
a neighbor having a lower degree.

A careful analysis of the $O(\sqrt{n})$ update time algorithm 
suggests that a {\em finer} partition of vertices may
help in achieving a better update time. This leads to our final algorithm which 
achieves expected amortized $O(\log n)$ time per update. 
More specifically, our algorithm maintains an invariant that can be informally 
summarized as follows.

{\em Each vertex tries 
to rise to a level higher than its current level, if, upon reaching that level, 
there are sufficiently large number of edges incident on it from lower levels. 
Once a vertex reaches a new level, it selects a 
random edge from this set and makes it matched. }

Note that we say that {\it "a vertex rises"} to indicate that the vertex moves to a higher level
and {\it "a vertex falls"} to indicate that the vertex moves to a lower level.
A vertex may also fall to a lower level if the number of edges incident to it 
decreases. But a vertex only uses its neighborhood information to decide 
whether to move to higher or lower level. %We will see that 
Overall, the vertices use
their local information to reach a global equilibrium state in which 
after each update each vertex is at the {\em right} level having
no incentive to either move above or below its current level.

\subsection{Organization of the paper}
For a gentle exposition of the ideas and techniques, we first 
describe a simple but less efficient fully dynamic algorithm for maximal 
matching in Section 4. We present our final fully dynamic algorithm which achieves expected amortized $O(\log n)$ time per update
in Section 5. In Section 6, we illustrate an example of a dynamic graph that establishes the tightness of the approximation factor guaranteed by our algorithm. In the following section, we describe notations and elementary results from the probability theory that we shall use. 
%This algorithm maintains a 2-level partition of the 
%vertices and achieves expected amortized $O(\sqrt{n})$ time per update. 
%In section 3, we present our final fully dynamic algorithm which has 
%$O(\log n)$ levels and achieves expected amortized $O(\log n)$ time per update 
%(Theorem \ref{theorem41}). 
%----------2016-edit-begins
%Though our algorithms use randomization very
%crucially to achieve efficiency, we show in Section 5 that the number of 
%random bits required by our algorithm is polylogarithmic.
%----------2016-edit-ends

%All logarithms in this paper are with base 2 unless mentioned otherwise. 
%During the analysis of algorithms, we shall use the terminology 
%{\it very high probability} for those events whose probability is
%$1-n^c$ for some positive constant $c$.%, i.e., inverse polynomial.

\section{Preliminaries}
We shall use ${\cal M}$ to denote the matching maintained by our algorithm at any stage. Our algorithms maintain a partition of the set
of vertices among various levels. We shall use $\LL(u)$ to denote the level of 
a vertex $u$. We define $\LL(u,v)$ for an edge $(u,v)$ as $\max(\LL(u),\LL(v))$.
Our algorithms will ensure that both the endpoints of each matched edge in ${\cal M}$ are present at the same level.  
%So we also keep the information about the level of each matched edge.
So the matching ${\cal M}$ maintained by our algorithms will be a set of tuples as follows. 
\[ {\cal M} = \{(u,v,\ell)~|~ u ~\mbox{is matched with}~v~\mbox{at level}~ \ell\}\]

The analysis of our algorithms will use a basic result about asymmetric random walk as follows.
\subsubsection*{Asymmetric random walk on a line}
Consider a particle performing a discrete random walk on a line. In each step, it moves one unit to the right
with probability $p$ or to the left with probability $q=1-p$. Each move is independent of the moves made in the past. 
The following lemma holds if $p>q$.
\begin{lemma}
%-2Aug2016-begin
Suppose the random walk starts at location $\ell$ units to the right of the origin. Then the probability that it ever reaches origin 
within $L$ steps for any given $L$ is less than $\left(\frac{q}{p}\right)^{\ell}$.
%-2-Aug2016-end
%Suppose the random walk starts at location $\ell$ units to the right of the origin. Then the probability that it ever reaches origin 
%within $\le L$ steps is at most
%\[ \frac{1-(p/q)^L}{1-(p/q)^{L+\ell}} < \left(\frac{q}{p}\right)^{\ell}  \]
\label{lemma:asymmetric-random-walk}
\end{lemma}
The proof of Lemma \ref{lemma:asymmetric-random-walk} is sketched in Appendix.
During the analysis of algorithms, we shall use the terminology 
{\it very high probability} for those events whose probability is
$1-n^c$ for some positive constant $c$.

\section{Fully dynamic algorithm with expected amortized $O(\sqrt{n})$ time per update}

The algorithm maintains a partition of the set
of vertices into two levels~- 0 and 1.
%We shall use $\LL(u)$ to denote the level of
%a vertex $u$. We define $\LL(u,v)$ for an edge $(u,v)$ as
%$\max(\LL(u),\LL(v))$.
We now introduce the concept of {\em ownership} of the edges.
Each edge present in the graph will be owned by one or both of its endpoints
as follows. If both the endpoints of an edge are at level $0$, then
it is owned by both of them. Otherwise it will be owned by exactly that
endpoint which lies at higher level. If both the endpoints are at level 1,
the tie will be broken suitably by the algorithm.
As the algorithm proceeds, the vertices will make transition from one level to
another and the ownership of edges will also change accordingly. Let
$\OO_u$ denote the set of edges owned by $u$ at any moment of time. Each vertex
$u\in V$ will keep the set $\OO_u$ in a dynamic hash table \cite{cuckoo} so
that each search or deletion operation on $\OO_u$ can be performed in worst case
$O(1)$ time and each insertion operation can be performed in expected
$O(1)$ time. This hash table is also suitably augmented with a linked
list storing $\OO_u$ so that we can retrieve all edges of set $\OO_u$ in $O(|\OO_u|)$ time.

The algorithm maintains the following three invariants after each update.
\begin{enumerate}

\item Every vertex at level 1 is matched. Every free vertex at level 0
has all its neighbors matched.

\item Every vertex at level 0 owns less than $\sqrt{n}$ edges at any moment
of time.

\item Both the endpoints of every matched edge are at the same level.
\end{enumerate}

The first invariant implies that the matching $\MM$ maintained is maximal at
each stage. A vertex $u$ is said to be a {\em dirty} vertex at a moment
if at least one of its invariants does not hold. In order to restore the
invariants, each dirty vertex might make transition to some
new level and do some processing. This processing involves owning
or disowning some edges depending upon whether the level of the vertex has
risen or fallen. Thereafter, the vertex will execute \textsc{random-settle} or
\textsc{naive-settle} to {\em settle down} at its new level. The pseudocode
of our algorithms for handling insertion and deletion of an edge is given in
Figure~\ref{Figure1} and Figure~\ref{Figure2}.

\subsubsection*{Handling insertion of an edge}

\begin{figure}[hpt]
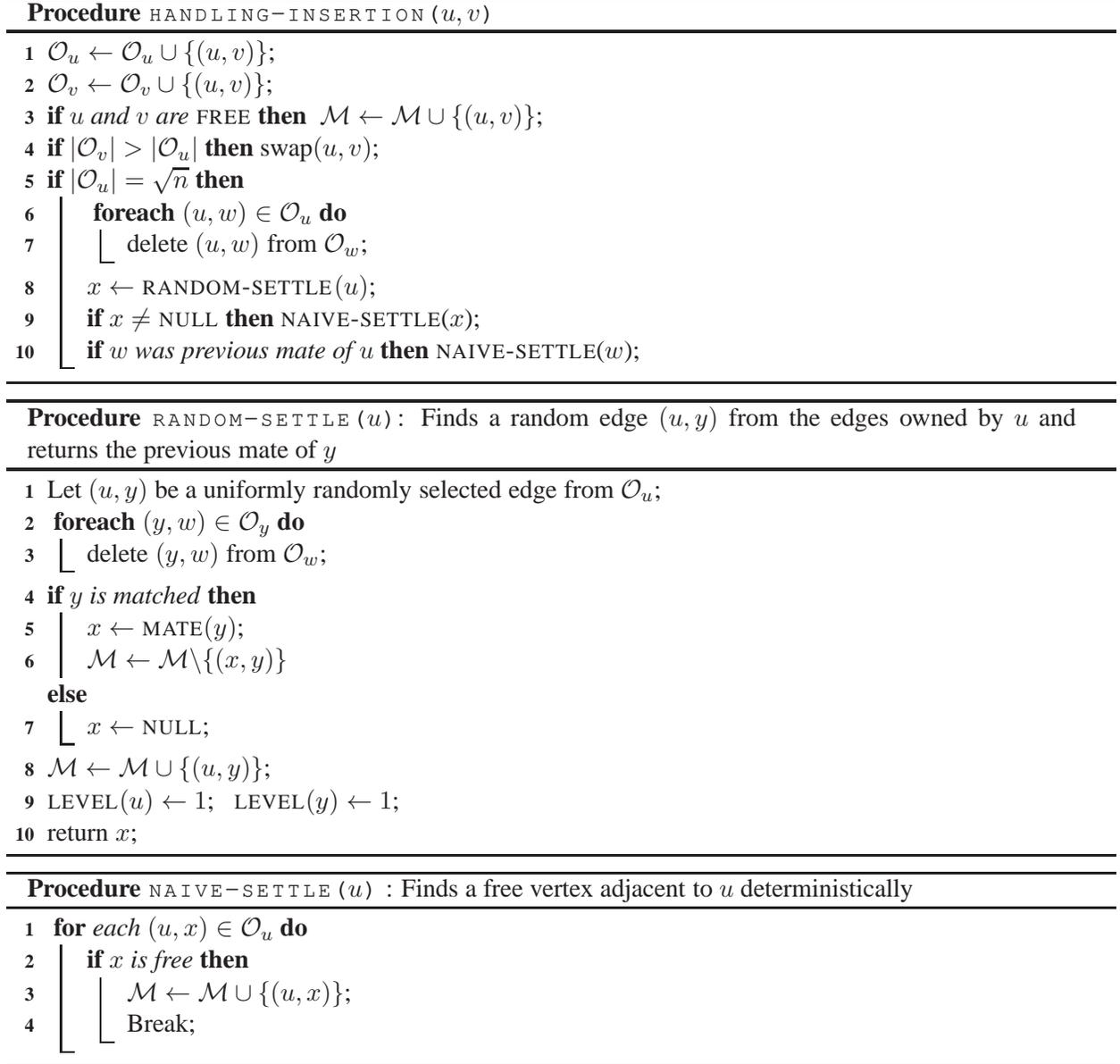

    \begin{procedure}[H]%
      \input{insertionhandle.tex}%
    \end{procedure}%

    \begin{procedure}[H]%
      \input{randomsettle.tex}%
    \end{procedure}%

    \begin{procedure}[H]%
      \input{naivesettle.tex}%
    \end{procedure}%

  \caption{Procedure for handling insertion of an edge $(u,v)$ where $\LL(u)=\LL(v)=0$.}
  \label{Figure1}
 \end{figure}

Let $(u,v)$ be the edge being inserted. If either $u$ or $v$ are at level 1,
there is no violation of any invariant. So the only processing that needs to be
done is to assign $(u,v)$ to $\OO_u$ if $\LL(u)=1$, and to $\OO_v$ otherwise.
This takes $O(1)$ time. However, if both $u$ and $v$ are at level 0, then
we execute \textsc{handling-insertion} procedure which does the
following (see Figure~\ref{Figure1}).

Both $u$ and $v$ become the owner of the edge $(u,v)$.
If $u$ and $v$ are free, then the insertion of $(u,v)$ has violated the
first invariant for $u$ as well as $v$. We restore it by adding
$(u,v)$ to $\MM$. Note that the insertion of $(u,v)$ also leads to increase of
$|\OO_u|$ and $|\OO_v|$ by one. We process that vertex from $\{u,v\}$
which owns larger number of edges; let $u$ be that vertex. If
$|\OO_u|=\sqrt{n}$, then Invariant 2 has got violated.
We execute \textsc{random-settle}($u$); as a result, $u$ moves to level 1 and
gets matched to some vertex, say $y$, selected randomly uniformly from
$\OO_u$. Vertex $y$ moves to level 1 to satisfy Invariant 3.
If $w$ and $x$ were respectively the earlier
mates of $u$ and $y$ at level 0, then the matching of $u$ with
$y$ has rendered $w$ and $x$ free.
So to restore Invariant 1, we execute \textsc{naive-settle}($w$)  and
\textsc{naive-settle}($x$). This finishes the processing of insertion of
$(u,v)$. Note that when $u$ rises to level 1, $|\OO_v|$ remains unchanged.
Since all the invariants for $v$ were satisfied before the current edge
update, it follows that the second invariant for $v$ still remains valid.

%%%%%-----------------Handling deletion of an edge---------------------%%%%%%
\subsubsection*{Handling deletion of an edge}
Let $(u,v)$ be an edge that is deleted. If $(u,v)\notin \MM$, all the
invariants are still valid. So let us consider the nontrivial case
when $(u,v)\in \MM$. In this case, the deletion of $(u,v)$ has made $u$ and
$v$ free. Therefore, potentially the first invariant might have got violated
for $u$ and $v$, making them dirty. We do the following processing in this
case.

If edge $(u,v)$ was at level 0, then following the deletion of $(u,v)$,
vertex $u$ executes \textsc{naive-settle}($u$), and then vertex $v$
executes \textsc{naive-settle}($v$). This restores the first invariant
and the vertices $u$ and $v$ are {\em clean} again. If edge $(u,v)$ was at
level 1, then $u$ is processed using the procedure shown in Figure
\ref{Figure2} which does the following ($v$ is processed similarly).
\begin{figure}[hpt]
\begin{procedure}[H]%
  \nl \ForEach{$(u,w)\in \OO_u$ ~{\bf and}~ $\LL(w)=1$}{
  \nl move $(u,w)$ from $\OO_u$ to $\OO_w$\;}
  \nl \eIf{$|\OO_u|\ge \sqrt{n}$}{
  \nl  $x\leftarrow \textsc{random-settle}(u)$\;
   \nl \lIf{$x\not= \textsc{null}$}\textsc{naive-settle}($x$)\;}{
   \nl $\LL(u) \leftarrow 0$\;
   \nl \ForEach{$(u,w)\in \OO_u$ ~{\bf and}~ $\LL(w)=0$}{
  \nl add $(u,w)$ to $\OO_w$\;}
  \nl  \textsc{naive-settle}($u$)\;
  \nl  \ForEach{$(u,w)\in \OO_u$}{
  \nl    \If{$|\OO_w| = \sqrt{n}$}{
   \nl     $x\leftarrow \textsc{random-settle}(w)$\;
   \nl     \lIf{$x\not= \textsc{null}$}\textsc{naive-settle}($x$)\;
      }
    }
  }
\caption{\textsc{handling-deletion}($u$,$v$)}
\end{procedure}
\caption{Procedure for processing $u$ when $(u,v)\in{\cal M}$ is deleted and
$\LL(u)$=$\LL(v)$=1.}
\label{Figure2}
\end{figure}

First, $u$ disowns all its edges whose other
endpoint is at level 1. If $|\OO_u|$ is still greater than
or equal to $\sqrt{n}$, then $u$ stays at level 1 and executes
{\textsc{random-settle}($u$)}.
If $|\OO_u|$ is less than $\sqrt{n}$, $u$ moves to level 0 and executes
{\textsc{naive-settle}($u$)}. Note that the transition of $u$ from level 1
to 0 leads to an increase in the number of edges owned by each of its
neighbors at level 0. The second invariant
for each such neighbor, say $w$, may get violated if $|\OO_w|=\sqrt{n}$,
making $w$ dirty. So we scan each neighbor of $u$ sequentially and
for each dirty neighbor $w$ (that is, $|\OO_w|= \sqrt{n}$),
we execute {\textsc{random-settle}($w$)} to
restore the second invariant. This finishes the processing of deletion of
$(u,v)$.

It can be observed that, unlike insertion of an edge, the deletion of an edge
may lead to creation of a large number of dirty vertices. This may happen if the
deleted edge is a matched edge at level 1 and at least one of its endpoints
move to level 0. \\

%
%%%%--------------------  Analysis of the algorithm  ---------------------%%%%%
\subsection{Analysis of the algorithm}
While processing the sequence of insertions and deletions of edges, an
edge may become matched or unmatched at different update steps.
We analyze the algorithm using the concept of
{\em epochs}, which we explain as follows.

\begin{definition}
At any time $t$, let $(u,v)$ be any edge in $\MM$. Then the {\bf epoch}
defined by $(u,v)$ at time $t$ is the maximal continuous time period
containing $t$ during which it remains in $\MM$. An epoch is said to
belong to level 0 or 1 depending upon the level of the matched edge that
defines the epoch.
\label{definition31}
\end{definition}

The entire life span of an edge $(u,v)$ consists of a sequence of
epochs of $(u,v)$ separated by the continuous periods when
$(u,v)$ is unmatched.
It follows from the algorithm that any edge update that does not change the
matching is processed in $O(1)$ time. An edge update that changes the matching
results in the start of new epoch(s) or the termination of some existing
epoch(s).
For the sake of analysis, we will redistribute the computation performed
at any update step $t$ among the epochs that are created or terminated at step
$t$. More specifically, let epoch of $(u_1,v_1), (u_2,v_2), \dots , (u_l,v_l)$
be created and epochs of $(w_1,x_1), (w_2,x_2), \dots, (w_k,x_k)$ be
terminated at step $t$. We will redistribute total computation performed
at step $t$ in such a way that:\\
\begin{tabular}{llll}
Total computation performed at step $t$ &=& $\sum_{i=1}^l$ computation associated with the start of epoch $(u_i,v_i)$ +\\
      &&$\sum_{i=1}^k$ computation associated with the termination of epoch $(w_i,x_i)$\\
\end{tabular}

Now, we shall analyze the computation involved in each procedure of our
algorithm and distribute it suitably among various epochs.

\begin{enumerate}

\item {\em \textsc{naive-settle}$(u)$}\\
Observe that whenever the procedure \textsc{naive-settle}$(u)$ is carried out,
$u$ is present at level 0, and hence $|{\cal O}_u|< \sqrt{n}$.
The procedure \textsc{naive-settle}$(u)$ searches for a free neighbor of $u$
by scanning $\OO_u$. Hence, the time complexity of \textsc{naive-settle}$(u)$
is $O(|{\cal O}_u|)$ = $O(\sqrt{n})$. Furthermore, this procedure is called
whenever $u$ loses its mate, say $v$. So we can associate the
computation cost of \textsc{naive-settle}$(u)$ with the termination of the
previous epoch $(u,v)$.

\item {\em \textsc{random-settle}$(u)$}\\
Observe that whenever {\em \textsc{random-settle}$(u)$} is invoked,
$u$ owns at least $\sqrt{n}$ edges incident from level 0, and
hence $|\OO_u|\ge \sqrt{n}$.
During \textsc{random-settle}$(u)$, $u$  finds a random mate
from level 0. This is done by selecting a random number $r \in [1,|\OO_u|]$,
and then picking the $r^{\tiny{th}}$ edge, say $(u,y)$, from the linked list
storing $\OO_u$. This takes $O(|\OO_u|)$ time.
Vertex $u$ then {\em pulls} $y$ to level 1 to satisfy the third invariant.
In this process, $y$ becomes the sole owner of all those edges whose
other endpoint is at level 0 (line 2,3). Since $y$ was the owner of at
most $\sqrt n$ edges, the total computation time involved in performing
this step is $O(\sqrt n)$. Other steps in \textsc{random-settle} can
be executed in $O(1)$ time. Hence the total computation time is
$O(|\OO_u| + \sqrt{n})$ which is $O(|\OO_u|)$ since $|\OO_u|\ge \sqrt{n}$.
We associate this computation time with the start of the epoch $(u,y)$ that gets
created at level 1.

\item {\em \textsc{handling-insertion}$(u,v)$}\\
This procedure takes $O(1)$ time unless one of the endpoints of $(u,v)$
starts owning $\sqrt{n}$ edges. In that case, the procedure invokes
\textsc{random-settle} (line 8) and  \textsc{naive-settle} (line 9,10).
We have already distributed the time taken in these procedures
to the respective epochs that get created.
Excluding these tasks, the only computation performed in this
procedure is in the \textsc{for} loop. The purpose of this loop is to
make $u$ the sole owner of all its edges incident from level 0.
Since $u$ owns $\sqrt n$ edges from level 0, the total
computation time involved in performing this step is $O(\sqrt n)$.
%
%parts, we will now try to associate the other part of
%this procedure to some epoch. In this procedure,
%a vertex $u$ executes \textsc{random-settle}$(u)$.
%In a preparation for this, it becomes the sole owner of its edges at level 0.
%Since $u$ was the owner of $\sqrt n$ edges, the total
%computation time involved in performing this step is $O(\sqrt n)$.
We associate this computation time with the start of the epoch created by $u$
at level 1.

\item {\em \textsc{handling-deletion}$(u,v)$}\\
Procedure \textsc{handling-deletion}$(u,v)$ is carried out when the matched
edge $(u,v)$ at level 1 gets deleted. In addition to invoking
\textsc{random-settle} and \textsc{naive-settle} procedures whose computation
cost is already assigned to respective epochs, this procedure scans the
list $\OO_u$ at most twice. Notice that $|\OO_u|$ can be $\Theta(n)$.
We associate this computation time of $O(n)$ with the termination of the epoch $(u,v)$.
\end{enumerate}

Excluding the updates that cause the start and
termination of an epoch of $(u,v)$, every other edge update on $u$ and $v$
during the epoch is handled in just $O(1)$ time. Therefore, we shall
focus only on the amount of computation associated with the
start and termination of an epoch. Let us now analyze the computation time
associated with the epoch at level 0 and level 1.

\begin{itemize}
\item Epoch at level 0 \\
As discussed above, it is only the procedure \textsc{naive-settle} whose
computation time is associated with an epoch at level 0. This procedure
takes $O(\sqrt{n})$ time. Hence the computation time associated with
an epoch at level 0 is $O(\sqrt{n})$.

\item Epoch at level 1\\
Consider an epoch at level 1. There are two ways in which this epoch gets
created at level 1:
\begin{itemize}
\item In \textsc{handling-insertion}\\
An epoch of $(u,v)$ can be created during the procedure
\textsc{handling-insertion}($u,v$). In this case, the computation time
associated with the start of the epoch of $(u,v)$ is
the computation time incurred in executing the procedure
\textsc{handling-insertion} and the procedure \textsc{random-settle}
which it invokes. It follows from the above discussion that the
computation cost associated with the epoch $(u,v)$ is $O(\sqrt{n} + |\OO_u|)$
which is $O(\sqrt{n})$ since $|\OO_u|=\sqrt{n}$ when we invoke
\textsc{handling-insertion}($u,v$).
\item In \textsc{handling-deletion}\\
Procedure \textsc{handling-deletion}$(u,v)$ invokes \textsc{random-settle}
at lines 4 and 12 to create new epochs at level 1. The execution of
\textsc{random-settle} at line 4 creates a new epoch for $u$ and its
computation time $O(|\OO_u|)$, which can be $\Theta(n)$, gets associated with
the start of the new epoch created by $u$. The execution of
\textsc{random-settle} at line 12 creates a new epoch for some vertex $w$
which is some neighbor of $u$. Note that $|\OO_w|=\sqrt{n}$.
Its computation time, which is $O(\sqrt{n})$, is associated
with the start of the epoch at level 1 created by $w$.

\end{itemize}

Now let us calculate the computation cost associated with an epoch, say of an
edge $(u,v)$, at level 1 when it terminates. It follows from the discussion above
that the only computation time associated with the termination of epoch $(u,v)$ is
the computation time of \textsc{handling-deletion} (excluding the time spent in
procedures \textsc{random-settle} and \textsc{naive-settle} that are already
associated with the start of their respective epochs).
This cost is at most $O(n)$.
\end{itemize}

From our analysis given above, it follows that the amount of computation time
associated with an epoch at level 0 is $O(\sqrt n)$ and the computation time
associated with an epoch at level 1 is $O(n)$.\\

An epoch of $(u,v)$ may either terminate (if $(u,v)$ is removed from the matching)
or remain {\em alive}, i.e., $(u,v)$ remain in the matching after the end of
all the updates. An epoch of $(u,v)$, ends because of
exactly one of the following causes.
\begin{enumerate}
\item[($i$)]

if $(u,v)$ is deleted from the graph.
\item[($ii$)]

$u$ (or $v$) get matched to some other vertex leaving its current mate free.
\end{enumerate}

An epoch will be called a {\em natural} epoch if it terminates due to
cause ($i$); otherwise it will be called an {\em induced} epoch. {\em Induced}
epoch can terminate prematurely since, unlike
natural epoch, the matched edge is not actually deleted from the graph
when an {\em induced} epoch terminates.

\begin{figure*}[!t]
\psfrag{Level2}{\textsc{level 1}}
\psfrag{Level1}{\textsc{level 0}}
\psfrag{natural}{\footnotesize{natural epoch}}
\psfrag{induced}{\footnotesize{induced epoch}}
\psfrag{time}{{\bf Time}}
\psfrag{e(u,w)}{\footnotesize{epoch of $(u,w)$}}
\psfrag{e(u,v)}{\footnotesize{epoch of $(u,v)$}}
\psfrag{e(v,x)}{\footnotesize{epoch of $(v,x)$}}
\centerline{\epsfysize=120pt \epsfbox{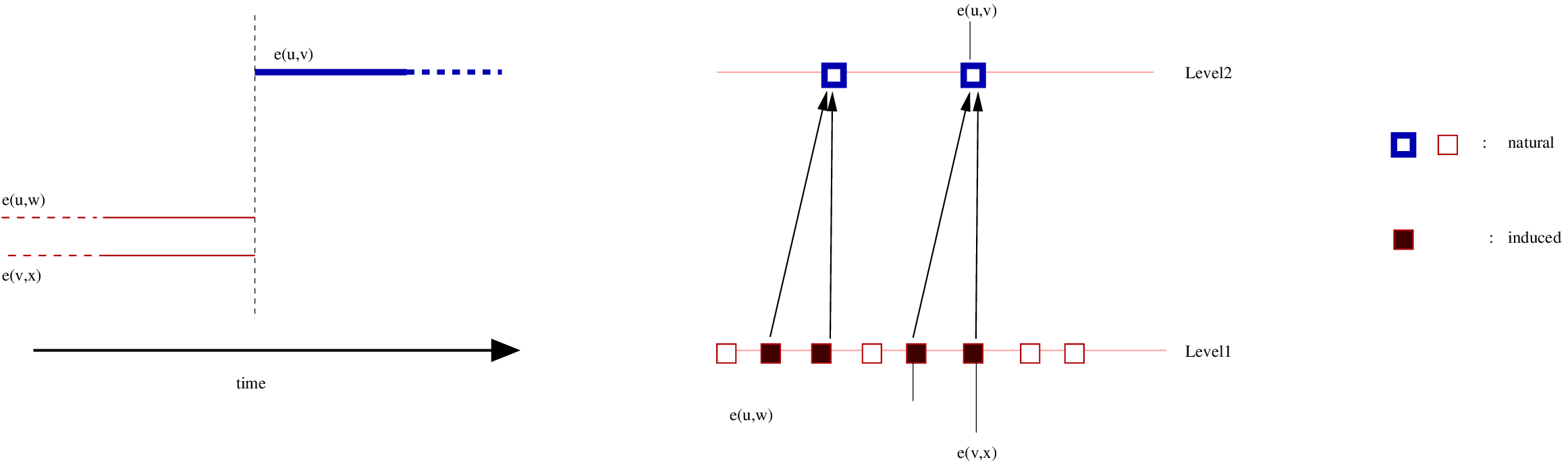}}
\caption{Epochs at level 0 and 1; the creation of an epoch at level 1 can
terminate at most two epochs at level 0.
}
\label{Figure3}
\end{figure*}

It follows from the algorithm described above that every epoch at level 1 is a
natural epoch whereas an epoch at level 0 can be natural or induced
depending on the cause of its termination. Furthermore,
each induced epoch at level 0 can be associated with a natural
epoch at level 1 whose creation led to the termination of the former.
In fact, there can be at most two induced epochs at level 0 which can be
associated with an epoch at level 1. It can be explained as follows (see Figure
\ref{Figure3}).

Consider an epoch at level 1 associated with an edge, say $(u,v)$. Suppose
it was created by vertex $u$.
If $u$ was already matched at level 0, let $w\not=v$ be its mate.
Similarly, if $v$ was also matched already, let $x\not=u$ be its current mate
at level 0. So matching $u$ to $v$ terminates the epoch of $(u,w)$ as well
as the epoch of edge $(v,x)$ at level 0. We {\em charge} the overall cost of
these two epochs to the epoch of $(u,v)$. We have seen that the computational
cost associated with an epoch at level 0 is $O(\sqrt n)$.
So the overall computation {\em charged} to an epoch of $(u,v)$ at level 1
is $O(n+2\sqrt n)$ which is $O(n)$.

\begin{lemma}
The computation charged to a natural epoch at level 1 is $O(n)$
and the computation charged to a natural epoch at level 0 is $O(\sqrt{n})$.
\label{lemma31}
\end{lemma}

In order to analyze our algorithm, we just need to get a bound on the
computation {\em charged} to all natural epochs that get terminated
during a sequence of updates or are alive at the end of the all the updates.
Let us first analyze the computation cost {\em charged} to all those epochs which
are alive at the end of $t$ updates. Consider an epoch of edge $(u,v)$ that is
alive at the end of $t$ updates. If this epoch is at level 0, the computation
cost associated with the start of this epoch is $O(1)$. If this epoch is at
level 1, then the computation time associated with the start of this epoch
is $O(|\OO_u|)$ and notice that $|\OO_u|\ge \sqrt{n}$. Note that there
can be at most two induced epochs at level 0 whose computation time, which is
$O(\sqrt{n})$, is also charged to the epoch of $(u,v)$. Hence the computation
charged to the live epoch of $(u,v)$ is $O(|\OO_u|)$. Observe that, at any
given moment of time, $\OO_u\cap \OO_w =\emptyset$ for any two vertices $u,w$
present at level 1. Hence the computation time charged to all live epochs at
the end of $t$ updates is of the order of $\sum_{u} |\OO_u| \le 2t = O(t)$.
So all we need is to analyse the computation charged to all natural epochs that
get terminated during the sequence of updates.

Let $t$ be the total number of updates. Each natural epoch at level 0 which
gets terminated can be assigned uniquely to the deletion of its matched edge.
Hence it follows from Lemma \ref{lemma31} that the computation
{\em charged} to all natural epochs terminated at level 0 during $t$
updates is $O(t\sqrt n)$. We shall now analyze the number of epochs terminated at level 1.
Our analysis will crucially exploit the following lemma.
\begin{lemma}
Suppose vertex $v$ creates an epoch at level $1$ during an update in the graph.
and let $O_v^{init}$ be the set of edges that $v$ owned at the time of the creation of this epoch. Then, for any arbitrary sequence $D$ of edge deletions of $O_v^{init}$, and for any $(v,w)\in O_v^{init}$
\[
\PP[\textsc{mate}(v)=w ~|~ D] ~~=~~  \frac{1}{|O_v^{init}|}
\]
\label{lemma:key-lemma-level-2-algorithm}
\end{lemma}
We first carry out the analysis for the high probability bound on the total update time taken by our
algorithm. Thereafter we carry out the analysis for the expected value of the total update time.

%Note that though an edge $e$ can be
%deleted and inserted multiple times in an update sequence, we shall treat
%each of these updates differently
%
%---------------------------------------------------------------------------------------------------%
                \subsection{High probability bound on the total update time}
%---------------------------------------------------------------------------------------------------%
The key idea of randomization is that once a vretex $v$ creates an epoch at level 1, there should be {\em many} edge deletions from
$\OO_v^{init}$ before the matched edge of $v$ is deleted. In order to quanitfy this key idea, we introduce the following definition that categorizes an epoch as good or bad.
\begin{definition} An epoch is said to be bad if it gets terminated naturally with in the deletion of the first $1/3$ edges that it
owned at the time of its creation. An epoch is said to be good if it is not bad.
\label{def:bad-epoch}
\end{definition}
\noindent
It follows from Definition \ref{def:bad-epoch} that a good epoch undergoes many edge deletions before getting terminated. So only the bad epochs are problematic. Now using Lemma \ref{lemma:key-lemma-level-2-algorithm}, we establish an upper bound on the probability of an epoch to be bad.
\begin{lemma}
Suppose vertex $v$ creates an epoch at level $1$ during the $k$th update for some $k\le t$.
Then this epoch is going to be bad with probability $1/3$ irrespective of the future updates in the graph and
the random bits picked during their processing.
\label{lemma:2-level-bad-epoch-prob}
\end{lemma}
\begin{proof}
Consider any sequence of updates in the graph following the creation of this epoch. This sequence defines
the sequence $D$ of edge deletions of $\OO_v^{init}$. The termination of this epoch is fully determined by the mate that $v$ picked
and this sequence $D$. This epoch will be bad if the mate of $v$ is among the endpoints of the first $1/3$ edges
in this sequence. Then, Lemma \ref{lemma:key-lemma-level-2-algorithm} implies that the mate of $v$ is equally likely to be the
endpoint of any edge in this sequence. %$\OO_v^{init}$.
So the probability of the epoch to be bad is $1/3$ \footnote{The analysis assumed that each edge from $\OO_v^{init}$ is going to be deleted sometime in future. If not, place all such edges arbitrarily at the end of the sequence $D$. In this case, %as the reader may also see, there are chances for the epoch to remain alive instead of getting terminated.
the probability of the epoch to be bad will be even less than $1/3$.}.
\end{proof}

For the time complexity analysis, we will show that the number of bad epochs may exceed the number of good
epochs by at most $O(\log n)$ with very high probability. Notice that the number of epochs created at level 1 is itself a random variable whose value may depend upon the updates in the graph as well as the random bits picked during their processing. However, as shown
by Lemma \ref{lemma:2-level-bad-epoch-prob}, each newly created epoch at level 1 will be bad with probability $1/3$
irrespective of the past epochs. The number of epochs created at level 1 during any $t$ updates is trivially $O(nt)$.
Therefore, the sequence of epochs at level 1 can be seen as an instance of the asymmetric random walk as follows. The walk starts at location
$2\log_2 n$ to the right of the origin. Each step of the walk is one unit to the right of the current location with probability $2/3$ or one unit to the left with probability $1/3$ independent of the past moves. We need to find the probability that the walk ever reaches the origin during any time
in the algorithm. It follows from Lemma \ref{lemma:asymmetric-random-walk} that the probability of this event is less than $1/n^2$. So the
following lemma holds immediately.

\begin{lemma}
During any sequence of $t$ updates, the number of bad epoch at level 1 can exceed the number of good epochs by $2 \log_2 n$ with probability at most $1/n^2$.
\label{lemma:level2-bad-epochs}
\end{lemma}

\noindent
As stated in Lemma \ref{lemma31}, each epoch at level 1 has a computation cost of $O(n)$ charged to it.
Let $t$ be the total number of updates in the graph. For each epoch at level 1, the number of owned edges at the time of its creation is at least $\sqrt{n}$. As a result the number of good epochs during $t$ updates is bounded by $4t/\sqrt{n}$ deterministically. So the computation cost of good epochs at level 1 is bounded by $O(t\sqrt{n})$. Lemma \ref{lemma:level2-bad-epochs} implies that the computation cost of all bad epochs at level 1 can exceed the computation cost of all good epochs at most by $cn \log n$ amount for some constant $c$ with probability $\ge 1-1/n^2$. So overall the cost of all epochs at level 1 is bounded by $O((t\sqrt{n} + n\log n)$ with high probability. The computation cost of all epochs at level 0 is bounded deterministically by $O(t\sqrt{n})$. Hence
the total computation time taken by our algorithm for any sequence of $t$ updates is $O(t\sqrt{n}+ n \log n)$ with high probability.

%---------------------------------------------------------------------------------------------------%
                \subsection{Expected value of the total update time}
%---------------------------------------------------------------------------------------------------%

Let $X_{v,i,k}$ be a random variable which is 1 if $v$ creates an epoch at level $i$ at update step $k$, otherwise it is 0. We denote this
epoch as \textsc{epoch}($v,i,k$).
Let $Z_{v,i,k}$ denote the number of edges from $\OO_v^{init}$ that are deleted during the epoch. (If \textsc{epoch}($v,i,k$) is not created,
$Z_{v,i,k}$ is defined as 0). Since each edge deletion at level 1 is uniquely associated to the epoch that owned it. Therefore,
$\sum_{v,k}Z_{v,1,k}\le t$. Hence,
\begin{equation}
\sum_{v,k}\EX[Z_{v,1,k}] \le t
\label{eq:Expected-edge-deletions-level-2}
\end{equation}

\noindent
We shall now derive a bound on the expected value of $Z_{v,1,k}$ in an alternate way.
\begin{lemma}
\label{lem:Conditional-expectation-of-Zv1k-level2}
 $\EX[Z_{v,1,k}] \ge \sqrt{n}/2 \cdot \PP[X_{v,1,k}=1]$.
\end{lemma}
\begin{proof}
We shall first find the expectation of $Z_{v,1,k}$ conditioned on the event that $v$ creates an epoch at level 1 during
$k$th update. That is, we shall find $\EX[Z_{v,1,k}|X_{v,1,k}=1]$.
Let ${\cal O}_v^{init}$ be the set of edges owned by $v$ at the moment of creation of \textsc{epoch}($v,1,k$), and let $D$ be the deletion sequence associated with $\OO_v^{init}$. It follows from Lemma \ref{lemma:key-lemma-level-2-algorithm} that the matched egde of $v$ is distributed uniformly over $\OO_v^{init}$. So $\EX[Z_{v,1,k}|X_{v,1,k}=1] = |\OO_v^{init}|/2 \ge \sqrt{n}/2$ since $|\OO_v^{init}|$ for an epoch at level $1$ is at least $\sqrt{n}$. Using conditional expectation, we get
\[ \EX[Z_{v,1,k}]  = \EX[Z_{v,1,k}|X_{v,1,k}=1] \cdot \PP[X_{v,1,k}=1] \ge \sqrt{n}/2 \cdot \PP[X_{v,1,k}=1]\]
\end{proof}
Notice that the computation cost of an epoch at level $1$ is at most $c n$ for some constant $c$.
So the expected value of the computation cost associated with all natural epochs that get terminated at level 1 during $t$ updates is
\begin{eqnarray*}
\sum_{v,k} cn\cdot \PP[X_{v,1,k}=1]
      & =   & 2 c\sqrt{n} \sum_{v,k} \sqrt{n}/2 \cdot \PP[X_{v,1,k}=1] \\
      & \le & 2 c\sqrt{n} \sum_{v,k} \EX[Z_{v,1,k}] ~~~~~~~\mbox{using Lemma \ref{lem:Conditional-expectation-of-Zv1k-level2}}\\
      & \le & 2 c\sqrt{n} t~~~~~~~\mbox{using Equation \ref{eq:Expected-edge-deletions-level-2}}
\end{eqnarray*}

\noindent
We can thus conclude with the following theorem.
\begin{theorem}
Starting with a graph on $n$ vertices and no edges, we can maintain maximal
matching for any sequence of $t$ updates in $O(t\sqrt{n})$ time in expectation
and $O(t\sqrt{n} + n\log n)$ with high probability.
\label{theorem31}
\end{theorem}

\subsection{On improving the update time beyond
$O(\sqrt{n})$}
In order to extend our 2-$\LL$ algorithm for getting a better update time,
it is worth exploring  the reason underlying $O(\sqrt{n})$ update time
guaranteed by our 2-$\LL$ algorithm. For this purpose, let us examine the
second invariant more carefully. Let $\alpha(n)$ be the threshold for the
maximum number of edges that a vertex at level 0 can own. Consider
an epoch at level 1 associated with some edge, say $(u,v)$.
The computation associated with this epoch is of the order of the number of
edges $u$ and $v$ own which can be $\Theta(n)$ in the worst case. However,
the expected duration of the epoch is of the order of the
minimum number of edges $u$ can own at the time of its creation, i.e.,
$\Theta(\alpha(n))$. Therefore, the expected amortized computation per edge
deletion for an epoch at level 1 is $O(n/\alpha(n))$. Balancing this with the
$\alpha (n)$ update time at level 0, yields $\alpha (n) = \sqrt{n}$.

In order to improve the running time of our algorithm, we need to decrease
the ratio between the maximum and the minimum number of edges a vertex can own
during an epoch at any level. It is this ratio that determines the
expected amortized time of an epoch.
This insight motivates us for having a finer partition of
vertices -- the number of levels should be increased to $O(\log n)$ instead of
just 2. When a vertex creates an epoch at level $i$, it will own at least
$2^{i}$ edges, and during the epoch it will be allowed to own at most
$2^{i+1}-1$ edges. As soon as it owns $2^{i+1}$ edges, it should
migrate to higher level. Notice that
the ratio of maximum to minimum edges owned by a vertex during an epoch
gets reduced from $\sqrt{n}$ to a constant.

We pursue the approach sketched above combined with some additional techniques
in the following section. This leads to a fully dynamic algorithm for maximal
matching which achieves expected amortized $O(\log n)$ update time per
edge insertion or deletion.

\label{sec:log}
\section{Fully dynamic algorithm with expected amortized $O(\log n)$ time per update}
This algorithm maintain a partition of vertices among various
levels.
%2016-begin
We describe the difference in this partition vis-a-vis 2-$\LL$ algorithm.
%2016-end
%We iterate the difference in the partition vis-a-vis 2-$\LL$ algorithm.

\begin{enumerate}

\item The fully dynamic algorithm maintains a partition of vertices among
$\lfloor \log_4 n\rfloor +2$ levels. The levels are numbered from $-1$ to
$\textsc{l}_0 = \lfloor \log_4 n \rfloor$.
%2016-begin
During the algorithm, when a vertex moves to level $i$, it owns at least $4^i$ edges.
%2016-end
%We will see that the algorithm moves a vertex to level $i$ if the
%vertex is the owner of at least $4^i$ edges at that moment.
So a vantage point is needed for a vertex that does not own any edge. As a result,
we introduce a level -1 that contains all the vertices that do not own any edge.

\item We use the notion of ownership of edges which is slightly different
from the one used in the 2-$\LL$ algorithm. In the 2-$\LL$ algorithm,
at level 0, both the endpoints of the edge are the owner of the edge.
Here, at every level, each edge is owned by exactly one of
its endpoints.
If the endpoints of the edge are at different levels, the edge
is owned by the endpoint that lies at the higher level.
%In particular, the endpoint at the higher level always owns the edge.
If the two endpoints are at the same level,
then the tie is broken appropriately by the algorithm.
\end{enumerate}

Like the 2-$\LL$ algorithm, each vertex $u$ will maintain a dynamic hash
table storing the edges $\OO_u$ owned by it.
In addition, the generalized fully dynamic algorithm will maintain  the
following data structure for each vertex $u$. For each $i\ge\LL(u)$,
let ${\cal E}_u^i$ be the set of all those edges incident on
$u$ from vertices at level $i$ that are not owned by $u$.
%For each vertex $u$ and level $i\ge \LL(u)$,
The set $\EE_u^i$ will be maintained in a dynamic hash table.
However, the onus of maintaining $\EE_u^i$ will not be on $u$.
For any edge $(u,v)\in \EE_u^i$, $v$ will be responsible for the maintenance
of $(u,v)$ in $\EE_u^i$ since $(u,v)\in \OO_v$.
%2016-begin
For example, suppose vertex $v$ moves to level $j$. If $j>\LL(u)$, then $v$
will remove $(u,v)$ from $\EE_u^i$ and insert it to $\EE_u^j$. Otherwise (
($j\le \LL(u)$),  $v$ will remove $(u,v)$ from $\EE_u^i$ and insert it to $\OO_u$.
%2016-end
%For example, when $v$ moves to a new level, say $j$,
%$v$ will remove $(u,v)$ from $\EE_u^i$ and insert it to $\EE_u^j$.

%%%%%%%-----------------------INVARIANTS---------------------------%%%%%%%%%
\subsection{Invariants and a basic subroutine used by the algorithm}

As can be seen from the 2-level algorithm, it is advantageous
for a vertex to get settled at a higher level once it owns a
{\em large} number of edges.
Pushing this idea still further, our fully dynamic algorithm will
allow a vertex to rise to a higher level if it can own {\em sufficiently
large} number of edges after moving there.
In order to formally define this approach, we introduce an important notation
here.

\fbox{\parbox{5.5in}{
For a vertex $v$ with $\LL(v)=i$,
\[
\phi_v(j) = \left\{ \begin{array}{ll}
|\OO_v| + \sum_{i\le k< j} |\EE_v^k| & \mbox{~if~} j>i \\
0 & \mbox{otherwise}
\end{array}
\right.
\]
}}

In other words, for any vertex $v$ at level $i$ and any $j>i$, $\phi_v(j)$
denote the number of edges which $v$ can own if $v$ rises to level $j$.
Our algorithm will be based on the following key idea.
If a vertex $v$ has $\phi_v(j) \ge 4^{j}$, then
$v$ would rise to the level $j$. In case, there are multiple levels to which
$v$ can rise, $v$ will rise to the highest such level. With
this key idea, we now describe the three invariants which our algorithm will
maintain.

\begin{figure}[!t]
\psfrag{l0}{$\textsc{l}_0$}
\psfrag{i}{\footnotesize{$i$}}
\psfrag{-1}{\footnotesize{$-1$}}
\psfrag{0}{\footnotesize{$0$}}
\psfrag{1}{\footnotesize{$1$}}
\psfrag{2}{\footnotesize{$2$}}
\psfrag{3}{\footnotesize{$3$}}
\psfrag{v}{\footnotesize{$v$}}
\psfrag{x}{\footnotesize{$x$}}
\psfrag{matched}{\footnotesize{matched edge}}
\psfrag{unmatched}{\footnotesize{unmatched edge}}
\psfrag{vdots}{$\vdots$}
\centerline{\epsfysize=140pt \epsfbox{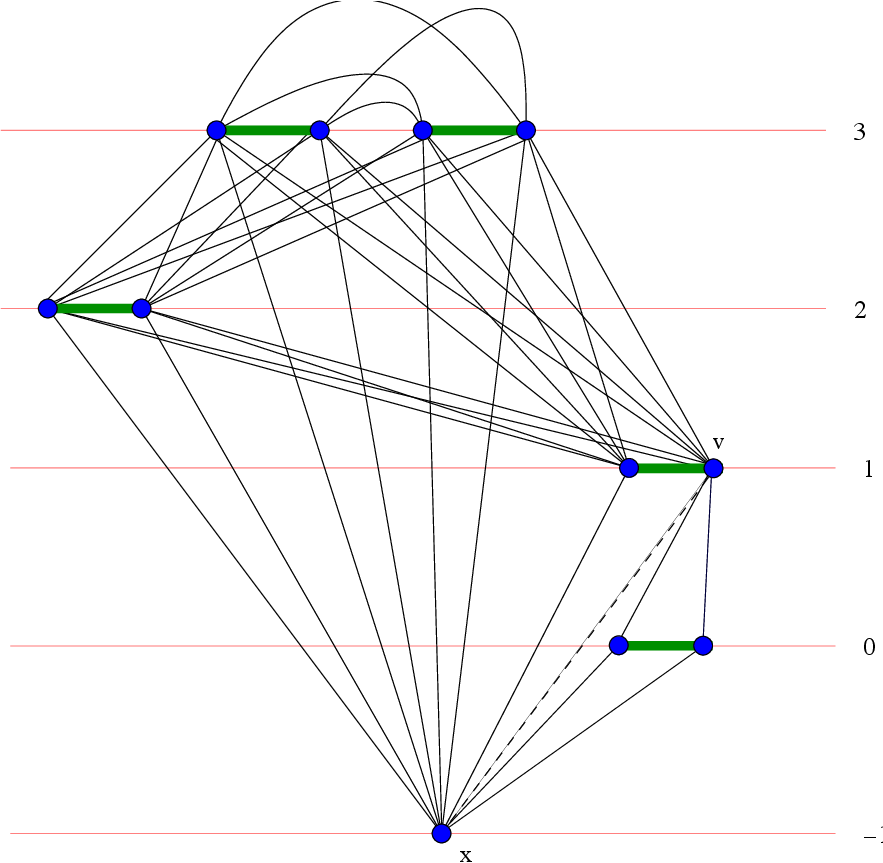}}
\caption{A snapshot of the algorithm on $K_9$: all vertices are matched( thick edges)
except vertex $x$ at level $-1$. %Vertex $v$ is the owner of just the edge $(v,x)$.
$\phi_v(2)=4 < 4^2$ and $\phi_v(3)=6 < 4^3$, so $v$ cannot rise
to a higher level. }
\label{Figure5}
\end{figure}

\begin{enumerate}
\item Every vertex at level $\ge 0$ is matched and every vertex at level $-1$ is
free.
\item For each vertex $v$ and for all
$j> \LL(v)$, $\phi_v(j) < 4^{j}$ holds true.

\item Both the endpoints of a matched edge are at the same level.
\end{enumerate}
It follows that the free vertices, if any, will be present at level $-1$ only.
Any vertex $v$ present at level $-1$ can not have any neighbor at level $-1$. Otherwise,
it would imply that $\phi_v(0) \ge 1 = 4^0$, violating the second invariant. Hence,
every neighbor of a free vertex must be matched. This implies that the
algorithm will always maintain a maximal matching.
Furthermore, the key idea of our algorithm is captured by the second invariant -- after processing
every update there is no vertex which fulfills the criteria of rising.
Figure \ref{Figure5} depicts a snapshot of the algorithm.

An edge update may lead to the violation of the invariants mentioned above and
the algorithm basically restores these invariants. This may involve rise or
fall of vertices between levels. Notice that the second invariant of a vertex is
influenced by the rise and fall of its neighbors. We now state
and prove two lemmas which quantify this influence more precisely.
\begin{lemma}
The rise of a vertex $v$ does not violate the second invariant for any of its neighbors.
\label{lemma41}
\end{lemma}
\begin{proof}
Consider any neighbor $u$ of $v$. Let $\LL(u) = k$.
Since the second invariant holds true for $u$ before the rise of $v$, so $\phi_u(i)<4^i$
for all $i>k$. It suffices if we can show that $\phi_u(i)$ does not increase
for any $i$ due to the rise of $v$. We show this as follows.

Let vertex $v$ rise from level $j$ to $\ell$. If $\ell\le k$, the edge
$(u,v)$ continues to be an element of $\OO_u$, and so there is no change in
$\phi_u(i)$ for any $i$. Let us consider the case when $\ell > k$. The
rise of $v$ from $j$ to $\ell$ causes removal of $(u,v)$ from $\OO_u$
(or $\EE_u^j$ if $j\ge k$) and insertion to $\EE_u^{\ell}$. As a result
$\phi_u(i)$ decreases by one for each $i$ in $[\max(j,k)+1$, $\ell]$, and
remains unchanged for all other values of $i$.
\end{proof}
\begin{lemma}
Suppose a vertex $v$ falls from level $j$ to $j-1$. As a result, for any neighbor $u$ of $v$, $\phi_u(i)$ increases by at most 1 for $i=j$ and remains unchanged
for all other values of $i$.
\label{lemma42}
\end{lemma}
\begin{proof}
Let $\LL(u) = k$.
In case $k\ge j$, there is no change in $\phi_u(i)$ for any $i$ due
to fall of $v$. So let us consider the case $j>k$. In this case, the fall of
$v$ from level $j$ to $j-1$ leads to the insertion of $(u,v)$ in
${\cal E}_u^{j-1}$ and deletion from ${\cal E}_u^{j}$. Consequently,
$\phi_u(i)$ increases by one only for $i=j$ and remains unchanged for all
other values of $i$.
\end{proof}

In order to detect any violation of the second invariant for
a vertex $v$ due to rise or fall of its neighbors, we shall
maintain $\{ \phi_v(i)| i\le \textsc{l}_0\}$ in an array $\phi_v[]$ of
size $\textsc{l}_0+2$. The updates on this data structure during
the algorithm will involve the following two types of operations.
\begin{itemize}
\item \textsc{decrement-$\phi$}($v,I$):~
This operation decrements $\phi_v(i)$ by one for all $i$ in interval $I$.
This operation will be executed when some neighbor of $v$ rises.
For example, suppose some neighbor of $v$ rises from level $j$ to $\ell$, then
$\phi_v(i)$ decreases by one for all $i$ in interval
$I=[\max(j,\LL(v))+1$, $\ell]$.

\item \textsc{increment-$\phi$}($v,i)$:~ this operation increases $\phi_v(i)$
by one. This operation will be executed when some neighbor of $v$ falls from
$i$ to $i-1$.
\end{itemize}
It can be seen that a single \textsc{decrement-$\phi$}$(v,I)$ operation takes
$O(|I|)$ time which is $O(\log n)$ in the worst case. On the other hand
any single \textsc{increment-$\phi$}$(v,i)$ operation takes $O(1)$ time.
However, since $\phi_v(i)$ is 0 initially and is non-negative always, we can
conclude the following.
\begin{lemma}
The computation cost of all \textsc{decrement-$\phi$}$()$ operations
over all vertices
is upper-bounded by the computation cost of all
\textsc{increment-$\phi$}$()$ operations over all vertices
during the algorithm.
\label{lemma43}
\end{lemma}
\begin{observation}
It follows from Lemma \ref{lemma43} that we just need to analyze the
computation involving all \textsc{increment}-$\phi()$ operations since the
computation involved in \textsc{decrement}-$\phi()$ operations is subsumed by
the former.
\label{observation41}
\end{observation}
If any invariant of a vertex, say $u$, gets violated, it might rise or fall,
though in some cases, it may still remain at the same level. However, in all
these cases, eventually the vertex $u$ will execute the procedure,
\textsc{generic-random-settle}, shown in Figure \ref{Figure6}. This
procedure is essentially a generalized version of \textsc{random-settle}$(u)$
which we used in the 2-level algorithm. \textsc{generic-random-settle}($u,i$)
starts with moving $u$ from its current level ($\LL(u)$) to level $i$. If
level $i$ is higher than the previous level of $u$, then $u$ performs the
following tasks. For each edge $(u,w)$ already owned by it, $u$
informs $w$ about its rise to level $i$ by updating $\EE_w^i$.
In addition $u$ acquires the ownership
of all the edges whose other endpoint lies at a level  $\in [\LL(u), i-1]$.
For each such edge $(u,w)$ that is now owned by $u$, we perform
\textsc{decrement-$\phi$}($w,[\LL(w)+1,i]$) to reflect that the edge is now
owned by vertex $u$ which has moved to level $i$. Henceforth, the procedure
then resembles \textsc{random-settle}. It finds a random edge $(u,v)$ from
$O_u$ and moves $v$ to level $i$. The procedure returns the previous mate of
$v$, if $v$ was matched. We can thus state the following lemma.

\begin{lemma}
Consider a vertex $u$ that executes \textsc{generic-random-settle}$(u,i)$ and
selects a mate $v$. Excluding the time spent in \textsc{decrement}-$\phi$
operations, the computation time of this procedure is of the order
of $|\OO_u| + |\OO_v|$ where $\OO_u$ and $\OO_v$ is the set of edges owned
by $u$ and $v$ just at the end of the procedure.
\label{lemma44}
\end{lemma}

%----------------------------------------------------------------------
%-----Pseudocode for Generic_Random_Settle(u,i)
%----------------------------------------------------------------------
%
\begin{figure}[hpt]
\begin{procedure}[H]
\nl\If(\tcp*[f]{\footnotesize\color{Brown}$u$ rises to level $i$}){$\LL(u)<i$}{
\nl\For(\tcp*[f]{\footnotesize\color{Brown}$u$ informs $w$ about its rise}){each $(u,w)\in \OO_u$}{
\nl     transfer $(u,w)$ from $\EE_w^{\LL(u)}$ to $\EE_w^i$\;
\nl     \textsc{decrement-$\phi$}($w,[\LL(u)+1,i]$)\;
   }
\nl\For(\tcp*[f]{\footnotesize\color{Brown}$u$ gains ownership of some more edges}){each $j=\LL(u)$ to $i-1$}{
\nl     \For{each $(u,w)\in \EE_u^j$}{
\nl         transfer $(u,w)$ from $\EE_u^j$ to $\EE_w^i$\;
\nl          transfer $(u,w)$ from $\OO_w$ to $\OO_u$\;
\nl          \textsc{decrement-$\phi$}($w,[j+1,i]$)\;
   }
}
\nl \lForEach{$j=\LL(u)+1$ to $i$}{$\phi_u(j) \leftarrow 0$\;}
\nl $\LL(u)\leftarrow i$\;
}

\nl Let $(u,v)$ be a uniformly randomly selected edge from $\OO_u$\;
\nl\eIf{$v$ is matched}{\nl $x\leftarrow \textsc{mate}(v)$\;
 \nl                    $\MM \leftarrow \MM \backslash \{(v,x)\}$\;}{
 \nl      $x \leftarrow \textsc{null}$\;}

\nl\For(\tcp*[f]{\footnotesize\color{Brown}$v$ informs $w$ about its rise}){each $(v,w)\in \OO_v$}{
\nl      transfer $(v,w)$ from $\EE_w^{\LL(v)}$ to $\EE_w^i$\;
\nl      \textsc{decrement-$\phi$}($w,[\LL(v)+1,i]$)\;
   }

\nl\For(\tcp*[f]{\footnotesize\color{Brown}$v$ gains ownership of some more edges}){each $j=\LL(v)$ to $i-1$}{
\nl     \For{each $(v,w)\in \EE_v^j$}{
\nl          transfer $(v,w)$ from $\EE_v^j$ to $\EE_w^i$\;
\nl          transfer $(v,w)$ from $\OO_w$ to $\OO_v$\;
\nl          \textsc{decrement-$\phi$}($w,[j+1,i]$)\;
     }
}
\nl $\MM \leftarrow \MM \cup \{(u,v)\}$\;
\nl \lForEach{$j=\LL(v)+1$ to $i$}{$\phi_v(j) \leftarrow 0$\;}
\nl $\LL(v)\leftarrow i$ \tcc*[r]{\footnotesize\color{Brown}{$v$ rises to level $i$}}
\nl return $x$;
\caption{\textsc{generic-random-settle}($u,i$)}
\end{procedure}
\caption{Procedure used by a free vertex $u$ to {\em settle} at \LL~$i$.}
\label{Figure6}
\end{figure}
%

%
%
%%%%%%%------------------------HANDLING EDGE UPDATES-----------------
%
\subsection{Handling edge updates by the fully dynamic algorithm}
Our fully dynamic algorithm will employ a generic procedure called
\textsc{process-free-vertices}(). The input to this procedure is a sequence
$S$ consisting of ordered pairs of the form $(x,k)$ where $x$ is a free
vertex at level $k\ge 0$. Observe that the presence of free vertices at level
$\ge 0$ implies that matching $\MM$ is not necessarily
maximal.
In order to preserve maximality of matching,
the procedure \textsc{process-free-vertices} restores the invariants of each
such free vertex till $S$ becomes empty.
We now describe our fully dynamic algorithm.
\subsubsection*{Handling deletion of an edge}

Consider deletion of an edge, say $(u,v)$. For each $j> \max(\LL(u),\LL(v))$,
we decrement $\phi_u(j)$ and $\phi_v(j)$ by one. If $(u,v)$ is an
unmatched edge, no invariant gets violated. So we only
delete the edge $(u,v)$ from the data structures of $u$ and
$v$. Otherwise, let $k=\LL(u)=\LL(v)$. We execute the Procedure
\textsc{process-free-vertices}($\langle (u,k),(v,k)\rangle$).
\subsubsection*{Handling insertion of an edge}
Consider insertion of an edge, say $(u,v)$. Without loss of generality,
assume that initially $u$ was at the same level as $v$ or a higher level than
$v$. So we add $(u,v)$
to $\OO_u$ and $\EE_v^{\LL(u)}$. For each $j> \max(\LL(u),\LL(v))$, we
increment $\phi_u(j)$ and $\phi_v(j)$ by one.
We check if the second invariant
has got violated for either $u$ or $v$. This invariant may get violated
for $u$ (likewise for $v$) if there is any integer $i>\max(\LL(u),\LL(v))$,
such that $\phi_u(i)$ has become $4^i$ just after the insertion of edge $(u,v)$.
In case there are multiple such integers, let $i_{\max}$ be the largest
such integer. To restore the invariant, $u$ leaves its current mate,
say $w$, and rises to level $i_{\max}$.
We execute \textsc{generic-random-settle}$(u,i_{\max})$, and let $x$ be the
vertex returned by this procedure.
Let $j$ and $k$ be respectively the levels of $w$ and $x$.
Note that $x$ and $w$ are two free vertices now.
We execute \textsc{process-free-vertices}($\langle (x,k),(w,j)\rangle$).

If the insertion of edge $(u,v)$ violates the second invariant for both $u$
and $v$, we proceed as follows. Let $j$ be the highest level to which $u$ can
rise after the insertion of $(u,v)$, that is, $\phi_u(j) = 4^j$.
Similarly, let $\ell$ be the highest level to which $v$ may rise, that is,
$\phi_v(\ell) = 4^\ell$. If $j\ge \ell$, we allow
only $u$ to rise to level $j$; otherwise we allow only $v$ to rise to $\ell$.
Note that after $u$ moves to level $j$, edge $(u,v)$
becomes an element of $\EE_v^j$. So $\sum_{\LL(v)\le k< \ell} |\EE_v^k|$
decreases by 1. As a result,
$\phi_v(\ell)= |\OO_v| + \sum_{\LL(v)\le k< \ell} |\EE_v^k|$ also decreases by
1 and is now strictly less than $4^\ell$; thus the second invariant for $v$
is also restored.

\subsubsection{Description of Procedure \textsc{process-free-vertices}}
\label{procedure-process-free-vertices}
The procedure receives a sequence $S$ of ordered pairs $(x,i)$ such that
$x$ is a free vertex at level $i$. It processes the free vertices
in a decreasing order of their levels starting from $\textsc{L}_0$.
We give an overview of this processing at level $i$.
For a free vertex at level $i$, if it owns {\em sufficiently} large number
of edges, then it settles at level $i$ and gets matched by selecting
a random edge from the edges owned by it. Otherwise the vertex falls down by
one level. Notice that the fall of a vertex from level $i$ to $i-1$ may lead
to rise of some of its neighbors lying at level $<i$. However, as follows
from Lemma \ref{lemma42}, for each such vertex $v$, only $\phi_v(i)$ increases
by one and $\phi_v()$ value for all other levels remains unchanged.
So the second invariant may get violated only for $\phi_v(i)$.
This implies that $v$ will rise only to level $i$. After these
rising vertices move to level $i$ (by executing \textsc{generic-random-settle}),
 we move onto level $i-1$ and proceed
similarly. Overall, the entire process can be seen as a wave of free vertices
falling level by level. Eventually this wave of free vertices reaches level
$-1$ and fades away ensuring maximal matching. With this overview,
we now describe the procedure in more details and its complete pseudocode is
given in Figure \ref{Figure7}.

\begin{figure}[hpt]
%
%%%%%%%---------------------PROCEDURE Process-free-vertices-----
%
\begin{procedure}[H]
\nl\lFor{each $(x,i)\in S$}{\textsc{enqueue}($Q[i],x$)}\;
\nl\For{$i=\textsc{l}_0$ to 0}{
\nl   \While{($Q[i]$ is not \textsc{empty})}{
\nl      $v\leftarrow $ \textsc{dequeue}($Q[i]$)\;
\nl       \eIf(\tcp*[f]{\footnotesize\color{Brown}$v$ falls to $i-1$}){\textsc{falling}($v$)}{
\nl            $\LL(v) \leftarrow i-1$\;
\nl            \textsc{enqueue}($Q[i-1],v$)\;
\nl
\nl            \For{each $(u,v) \in \OO_v$}{
\nl               transfer $(u,v)$ from $\EE_u^{i}$ to $\EE_u^{i-1}$\;
\nl               \textsc{increment-$\phi$}($u,i)$\;
\nl               \textsc{increment-$\phi$}($v,i)$\;
\nl               \If(\tcp*[f]{\footnotesize\color{Brown}$u$ rises to $i$}){$\phi_u(i)\ge 4^i$}{
\nl                  $x \leftarrow $ \textsc{generic-random-settle}($u,i$)\;
\nl                  \If{$x\not= \textsc{null}$}{\nl $j \leftarrow \LL(x)$\;
\nl                  \textsc{enqueue}($Q[j],x$);}
                }
            }
\nl       }(\tcp*[f]{\footnotesize\color{Brown}$v$ settles at level $i$}){
\nl            $x \leftarrow $ \textsc{generic-random-settle}($v,i$)\;
\nl            \If{$x\not= \textsc{null}$}{\nl $j \leftarrow \LL(x)$\;
\nl            \textsc{enqueue}($Q[j],x$);}
      }
   }
}
\caption{\textsc{process-free-vertices}($S$)}
\end{procedure}

%
%%%%%%%----------------FUNCTION Falling()-----------------
%
\begin{function}[H]
\nl $i\leftarrow \LL(v)$\;
\nl \For(\tcp*[f]{\footnotesize\color{Brown}$v$ disowns all edges at level $i$}){each $(u,v) \in \OO_v$ such that \LL($u$) = $i$ }{
\nl transfer $(u,v)$ from ~$\OO_v$~ to ~$\OO_u$\;
\nl transfer $(u,v)$ from ~$\EE_u^i$~ to ~$\EE_v^i$\;
}
\nl \lIf{$|\OO_v|< 4^{i}$}{return \textsc{true}~~}\lElse{
 return \textsc{false}\;
}
\caption{\textsc{falling}($v$)}
\end{function}
\caption{Procedure for processing free vertices given as a sequence $S$ of
ordered pairs $(x,i)$ where $x$ is a free vertex at \LL ~$i$.}
\label{Figure7}
\end{figure}

The procedure uses an array $Q$ of size $\textsc{l}_0+2$, where
$Q[i]$ is a pointer to a queue (initially empty) corresponding to level $i$.
For each ordered pair $(x,k)\in S$, it inserts $x$ into queue $Q[k]$. The
procedure executes a {\textsc{for}} loop from $\textsc{l}_0$ down to 0 where the $i$th
iteration extracts and processes the vertices of queue $Q[i]$ one by one as
follows. Let $v$ be a vertex extracted from $Q[i]$. First we execute the
function \textsc{falling}($v$) which does the following. $v$ disowns all its
edges whose other endpoint lies at level $i$. If $v$ owns less than
$4^{i}$ edges then $v$ falls to level $i-1$, otherwise $v$ will
continue to stay at level $i$.
%2016-begin
The processing of the free vertex $v$ for each of these two cases is done as follows.
%2016-end
%Depending on the outcome of
%\textsc{falling}($v$), further processing is done as follows.

%\begin{center}
%\begin{figure}[h]
%\fbox{\parbox{6.2in}{
\begin{enumerate}
\item {\em $v$ has to stay at level $i$.}
\\
$v$ executes \textsc{generic-random-settle} and selects a random mate,
say $w$, from level $j<i$ (if $w$ is present in $Q[j]$ then it is removed from
it and is raised to level $i$). If $x$ was the previous mate
of $w$, then $x$ is a falling vertex. Vertex $x$ gets added to $Q[j]$.
This finishes the processing of $v$.
\item
%2016-begin
{\em $v$ has to fall.}
%2016-ends
%{\em $v$ owns less than $4^{i}$ edges and has to fall.}
\\
In this case, $v$ falls to
level $i-1$ and is inserted to $Q[i-1]$. At this stage, $\OO_v$ consists of
neighbors of $v$ from level $i-1$ or below. It follows from Lemma
\ref{lemma42} that the fall of $v$ from $i$ to $i-1$ leads to increase in
$\phi_u(i)$ by one for each neighbor $u$ of $v$ which is present at a level
lower than $i$.
Moreover, $\phi_v(i)$, that was 0 initially, has to be set to $|\OO_v|$.
So all the vertices of $\OO_v$ are scanned, and for each $(u,v)\in \OO_v$,
we increment $\phi_u(i)$ and $\phi_v(i)$ by 1.
In case $\phi_u(i)$ has become $4^i$, $u$ has to rise to level $i$ and is
processed as follows. $u$ executes \textsc{generic-random-settle}$(u,i)$ to
selects a random mate, say $w$, from level $j<i$. If $w$ was in $Q[j]$
then it is removed from it. If $x$ was the previous mate of $w$, then $x$ is
a falling vertex, and so it gets added to queue $Q[j]$.
\end{enumerate}
%\caption{Processing of a free vertex $v$ from $Q[i]$.}
%\label{figure:processing-of-free-vertex}
%\end{figure}
%}}
%\end{center}

\begin{remark}
Notice a stark similarity between the above procedure for handling a free vertex
and the procedure for handling a free vertex at level 1 in the 2-level algorithm.
\label{remark:similarity-with-2-levels}
\end{remark}

In case 1, $v$ remains at level $i$ and $w$ moves to the level $i$ from some
level $j<i$. This renders vertex $x$ (earlier mate of $w$) free and the
first invariant of $x$ is violated. So $x$ is added to the queue at level $j$.
The processing of $v$ does not change $\phi_u()$ for any neighbor $u$ of $v$.
Furthermore, the rise of $w$ to level $i$ does not lead to violation of
any invariant due to Lemma \ref{lemma41}.
In case 2, $v$ falls to level $i-1$ and as a result some vertices may rise
to level $i$. Each such rising vertex executes \textsc{generic-random-settle}.
As in case 1, the processing of these rising vertices may create some
free vertices only at level $<i$. We can thus state the following lemma.
\begin{lemma}
After $i$th iteration of the for loop of
\textsc{process-free-vertices}, the free vertices are present only in the
queues at level $<i$, and for all vertices not belonging to these queues
the three invariants holds.
\label{lemma45}
\end{lemma}

Lemma \ref{lemma45} establishes that after termination of
procedure \textsc{process-free-vertices}, there are no free vertices at level
$\ge 0$ and all the invariants get restored globally.

%\pagebreak
%%%%%%----------------ANALYSIS OF THE ALGORITHM----------------------
%

%2016-begin
%2016-ends
%\begin{remark}
%Compared to the 2-$\LL$ algorithm described in the previous section, there is
%the following difference regarding the ownership of edges in the algorithm
%described above. Though the ownership of an
%edge may change as the algorithm processes the updates, for any given level
%$i$, the ownership of edges remains disjoint. The following
%example may illustrate this fact.
%Consider the moment when a vertex $u$ at level $i$ selects its
%matched edge, say $(u,v)$. At this moment, let $(u,w)$ be some other edge
%owned by $u$. As the algorithm processes updates, vertex $w$ may move to some
%level higher than $i$. At that moment, the ownership of the edge $(u,w)$
%will be transferred to $w$. But, if ever in future and so long as
%$u$ remains at level $i$, $w$ returns to level $i$, the ownership of
%edge $(u,w)$ will switch to $u$ again. Thus the ownership of an edge may change
%but each edge is owned by exactly one endpoint at every moment.
%\label{disjoint-ownership}
%\end{remark}

\subsection{Analysis of the algorithm}
Processing the deletion or insertion of an edge $(u,v)$ begins with
decrementing or incrementing $\phi_u(i)$ and $\phi_v(i)$ for each level
$j> \max(\LL(u),\LL(v))$. Since there are $O(\log n)$ levels, the computation associated with this
task over any sequence of $t$ updates will be $O(t \log n)$.
This task may be followed by executing the procedure
\textsc{process-free-vertices} that restores the invariants and updates the matching accordingly.
The updates in the matching can be seen as creation of new epochs and termination of some of the
existing epochs.
%We would like to mention an important point
%here. Along with other processing, the execution of this procedure involves
%\textsc{increment}-$\phi()$ and \textsc{decrement}-$\phi()$ operations.
%However, as implied by Observation \ref{observation41},
%the computation involving \textsc{decrement}-$\phi()$ is subsumed by
%\textsc{increment}-$\phi()$ operations.
%Our analysis of the entire computation performed while processing
%a sequence of $t$ updates is along similar lines to the 2-$\LL$ algorithm.
Like 2-level algorithm, for the purpose of analysis, we visualize the entire algorithm as a sequence of creation and termination
of various epochs. Excluding the $O(t\log n)$ time for maintaining $\phi$,
the total computation performed by the algorithm can be
associated with all the epochs that get terminated and
those that remain alive at the end of the sequence of updates. Along exactly similar lines
as in 2-level algorithm, the computation associated with all the epochs that are alive
at the end of $t$ updates is $O(t)$ only. So we just need to focus on the epochs that get
terminated and the computation associated with each of them.

Let us first analyse the computation associated with an epoch of a matched edge $(u,v)$. Suppose this
%
%2016-end
%Our analysis of the entire computation performed while processing
%a sequence of $t$ updates is along similar lines to the 2-$\LL$ algorithm.
%We visualize the entire algorithm as a sequence of creation and termination
%of various epochs. All we need to do is to analyze the number of
%epochs created and terminated during the algorithm and computation associated
%with each epoch.
%
%Let us analyze an epoch of a matched edge $(u,v)$. Suppose this
%
epoch got created by vertex $v$ at level $j$. So $v$ would have executed
\textsc{generic-random-settle} and selected $u$ as a random mate from level
$<j$. Note that $v$ must be owning less than $4^{j+1}$ edges and
$u$ would be owning at most $4^j$ edges at that moment. This observation and
Lemma \ref{lemma44} imply that the
computation involved in the creation of the epoch is $O(4^j)$. Once the epoch is
created, any update pertaining to $u$ or $v$ will be performed in just $O(1)$
time until the epoch gets terminated. Let us analyze the computation performed
when the epoch gets terminated. At this moment either one or both $u$ and $v$
become free vertices. If $v$ becomes free, $v$ executes the following task
(see procedure \textsc{process-free-vertices} in Figure
\ref{Figure7}):
$v$ scans all edges owned by it, which is less than $4^{j+1}$, and disowns
those edges incident from vertices of level $j$. Thereafter, if $v$ still
owns at least $4^j$ edges, it settles at level $j$ and creates a new
epoch at level $j$. Otherwise, $v$ keeps falling one level
at a time. For a single fall of $v$ from level $i$ to $i-1$, the computation
performed involves the following tasks: scanning the edges owned by $v$,
disowning those incident from vertices at level $i$, incrementing $\phi_w$
values for each neighbor $w$ of $v$ lying at level less
than $i$, and updating $\phi_v(i)$ to $|\OO_v|$. All this computation is of
the order of the number of edges $v$ owns at level $i$ which is less than
$4^{i+1}$. Eventually either $v$ settles at some level $k\ge 0$ and
becomes part of a new epoch or it reaches level $-1$. The total computation
performed by $v$ is, therefore, of the order of
$\sum_{i=k}^{j}4^{i+1}= O(4^j)$. This entire computation involving $v$
(and $u$) in this process is associated with the
the epoch of $(u,v)$. Hence we can state the following Lemma.
%
%%%%%---------------Lemma for computation involved in an epoch---------%%%%%%
%
\begin{lemma}
For any $i\ge 0$, the computation associated with an epoch at level $i$ is
$O(4^i)$.
\label{lemma46}
\end{lemma}

%2016-begin
%Let us now analyze the number of epochs terminated during any sequence of $t$
%updates.
%2016-end
%2016-begin

An epoch corresponding to edge $(u,v)$ at level $i$ could be
terminated if the matched edge $(u,v)$ gets deleted. Such an epoch is called
a natural epoch. However, this epoch could be terminated due to one of the
following reasons also.
\begin{itemize}
\item $u$ (or $v$) get selected as a random mate by one of their neighbors
present at $\LL>i$.
\item $u$ (or $v$) starts owning $4^{i+1}$ or more edges.
\end{itemize}
Each of the above factors render the epoch to be an induced epoch.
For any level $i>0$, the creation of an epoch causes termination of at most
two epochs at levels $<i$. It can be explained as follows:
Consider an epoch at level $i$ associated with an edge, say $(u,v)$. Suppose
it was created by vertex $u$.
If $u$ was already matched at some level $j<i$, let $w\not=v$ be its mate.
Similarly, if $v$ was also matched already at some level $k<i$, let $x\not=u$ be its mate.
So matching $u$ to $v$ terminates the epoch of $(u,w)$ and
$(v,x)$ at level $j$ and $k$ respectively.  We can thus state the following lemma.
\begin{lemma}
Creation of an epoch at a level $i$ may cause termination of at most 2 epochs at level $<i$.
\label{lemma:atmost-2-induced-epochs}
\end{lemma}

%2016-end
%
%We shall assign the cost of each induced epoch to the epoch which led to the
%termination of the former. For achieving this objective,
%we now introduce the notion of computation {\em charged} to an epoch at
%any level $i$. Note that no epoch is created at level -1 as the vertices at
%level -1 are always free. If $i=0$, the computation {\em charged} to the epoch
%is the actual computation performed during the epoch which is $O(1)$.
%For any level $i>0$, the creation of an epoch causes termination of at most
%two epochs at levels $<i$. It can be explained as follows:
%Consider an epoch at level $i$ associated with an edge, say $(u,v)$. Suppose
%it was created by vertex $u$.
%If $u$ was already matched at level $j (j<i)$, let $w\not=v$ be its mate.
%Similarly, if $v$ was also matched already, let $x\not=u$ be its current mate
%at level $k$. So matching $u$ to $v$ terminates the epoch of $(u,w)$ and
%$(v,x)$ at level $j$ and $k$ respectively. We {\em charge} the overall cost of
%these two epochs to the epoch of $(u,v)$.

%------------------------------------------------------------------------------%
              \subsubsection{Analysing an epoch}
%------------------------------------------------------------------------------%

%Consider any arbitrary sequence of $t$ updates in the graph. While processing these updates for maintaining maximal
%matching, various epochs get created and terminated at various levels. In order to analyse the total time taken by the algorithm
%we shall analyse these epochs.

Consider an epoch created by a vertex $v$ at level $i$.
At the time of the creation of the epoch, let $\OO_v^{init}$ be the set of edges owned by $v$, and let $w=\textsc{mate}(v)$.
This epoch may terminate much before the deletion of $(v,w)$.
This happens when $v$ or $w$ moves to some level $>i$ before the deletion of $(v,w)$.
In order to analyse termination of an epoch, therefore, we associate an update sequence with it as follows. For each edge $(v,x)\in \OO_v^{init}$, we consider the first time in future that $x$ moves to some higher level
\footnote{vertex $x$ may move to level $>i$ (and down) multiple times while the algorithm processes a sequence of updates. However, it is only the first time (after the creation of the epoch) when $x$ moves to a level $>i$ that is relevant as far as the possibility of the termination of the epoch by the upward movement of $x$ is concerned.}. The {\em update label} associated with edge $(v,x)$ is defined as
\begin{quote}
if $x$ moves to a level $> i$ before its deletion then it is classified as {\em upward}
else it is {\em deletion}.
\end{quote}
Likewise, we also consider the first time in future that $v$ moves to a level
$>i$. If $v$ never moves to any level $>i$ in future, we just append $v$ at the end of all the updates associated with $\OO_v^{init}$.
The update sequence $U$ for the epoch is the sequence of these updates on the edges of $\OO_v^{init}$ and vertex $v$ arranged in the chronological order. Consider the following example. Suppose $\OO_v^{init}$ has 10 edges and let the corresponding neighbors of $v$ be $\{w_1,\ldots,w_{10}\}$. Let the updates in the chronological order be :
the deletion of $(v,w_4)$, upward movement of $w_1$, upward movement of $w_9$, the deletion of $(v,w_5)$, and so on.
The corresponding update sequence will be
\[
U~:~\langle ~\stackrel{\bullet}{w_4}, ~
          \stackrel{\uparrow}{w_1},~
          \stackrel{\uparrow}{w_9},~
          \stackrel{\bullet}{w_5},~
          \stackrel{\bullet}{w_8},~
          \stackrel{\bullet}{w_3},~
          \stackrel{\uparrow}{w_2},~
          \stackrel{\uparrow}{v},~
          \stackrel{\bullet}{w_7},~
          \stackrel{\uparrow}{w_{10}},~
          \stackrel{\bullet}{w_8}~
 \rangle
\]
\begin{observation}
If the update associated with (the owner) $v$ appears at $\ell$th location in $U$, then the epoch will terminate on or before
the $\ell$th update in $U$. Therefore, the updates at location $>\ell$ in $U$ will have no influence on the termination of the epoch.
\label{obs:role-of-v-in-epoch}
\end{observation}
Unlike the 2-level algorithm, the update sequence associated with an epoch is not uniquely defined by the sequence of updates in the graph after the creation of the epoch. Rather, it also depends upon the current matching as well as the random bits chosen by the algorithm while
processing the updates.  So there is a probability distribution defined over all possible update sequences that depends upon these two factors. Consequently, the analysis of an epoch in our final algorithm is more complex compared to the 2-level algorithm.
In particular, it is not obvious whether there is any dependence between the random mate picked by a vertex while creating an epoch and the sequence of updates associated with the epoch.
However, using an interesting non-trivial property of our algorithm, we will establish that there is no dependence between the two.
\begin{lemma}
Suppose a vertex $v$ creates an epoch and let $\OO_v^{init}$ be the set of its owned edges at the time of the creation of this epoch. Then, for any update sequence $U$ and for each $(v,w)\in \OO_v^{init}$,
\[
\PP[\textsc{mate}(v)=w~ |~ U] ~=~ \PP[\textsc{mate}(v)=w] ~=~\frac{1}{|\OO_v^{init}|}
\]
\label{lemma:key-lemma-generic-algo}
\end{lemma}
Lemma \ref{lemma:key-lemma-generic-algo} can be seen as a generalization of Lemma \ref{lemma:key-lemma-level-2-algorithm} that we stated for our 2-level algorithm. Its proof is given in Section \ref{section:proof-of-key-lemma-generic-algo}. The analysis of our algorithm will be critically dependent on this lemma. Using this lemma,
we shall first establish a high probability bound on the total update time of the algorithm to process a sequence of updates in the graph.

%---------------------------------------------------------------------------------------------------------------%
               \subsubsection{High probability bound on the total update time}
%---------------------------------------------------------------------------------------------------------------%
%A vertex $v$ may create an epoch at any level during the processing of an update by the generic algorithm.
%Let \textsc{epoch}$(v,i,k)$ denotes such an epoch created by vertex $v$ at level $i$
%while the algorithm processes $k$th update in the graph.
%2016-begin
Recall Definition \ref{def:bad-epoch} of a bad epoch. It can be observed from this definition that an induced epoch is always a good epoch.
%2016-end
Using Lemma \ref{lemma:key-lemma-generic-algo},
the lemma for the probability of a bad epoch extends seamlessly from 2-level algorithm to our final algorithm as follows.
\begin{lemma}
Suppose vertex $v$ creates an epoch at level $i$ while the algorithm processes $k$th update in the graph.
This epoch will be bad with probability {\underline{at most}} $1/3$ irrespective of the updates in the graph and
the random bits picked during their processing.
\label{lemma:generic-bad-epoch-prob}
\end{lemma}
\begin{proof}
The termination of the epoch is completely
determined by the mate that $v$ picks and the update sequence associated with this epoch.
%Commented on 2nd August
%Let $({\cal U},P)$ be the probability distribution of all update sequences associated with this epoch.
Consider any update sequence $U$ associated with this epoch.
%\in {\cal U}$.
It follows from Lemma \ref{lemma:key-lemma-generic-algo} that conditioned on $U$, the mate of $v$ is equally likely to be the endpoint of any edge in $\OO_v^{init}$. Now recall from Observation \ref{obs:role-of-v-in-epoch} that for the epoch to
be terminated naturally, the mate of $v$ must be among the endpoints of the
deleted edges that precede $v$ in $U$. We distinguish between the
following two cases.
\begin{enumerate}
\item[Case1.] There are less than $|\OO_v^{init}|/3$ edge deletions preceding $v$ in $U$.\\
The epoch will be bad only if the matched edge of $v$ is one of these edge deletions preceding $v$ in $U$. Since the number of these
edge deletions is less than  $|\OO_v^{init}|/3$, so using Lemma \ref{lemma:generic-bad-epoch-prob} the probability of the epoch to be bad is less than $1/3$.
\item[Case2.] There are at least $|\OO_v^{init}|/3$ edge deletions preceding $v$ in $U$.\\
The epoch will be bad if the matched edge of $v$ is one of the first $|\OO_v^{init}|/3$ edge deletions in $U$.
% If the matched edge happens to belong to the subsequent edge deletions in $S$, the epoch will be good.
From Lemma \ref{lemma:key-lemma-generic-algo}, the termination of the
epoch is equiprobable for any of the $\OO_v^{init}$'s,
so the probability that the epoch is bad in this case is exactly $1/3$.
\end{enumerate}
It follows that the epoch %of $v$ created at $i$th level during $k$th update
is going to be bad with probability $1/3$ for each possible update sequence $U$ associated with the epoch.
%
%$\PP[\textsc{epoch}(v,i,k) ~\mbox{is bad}~|~U]\le 1/3$ for every %Therefore,
%\[
%   \PP[\textsc{epoch}(v,i,k) ~\mbox{is bad}]
%   ~~=~~   \sum_{U \in {\cal U}} \PP[\textsc{epoch}(v,i,k) ~\mbox{is bad}~|~U] ~\cdot~ \PP[U]
%   ~~\le~~ \sum_{U \in {\cal U}} \frac{1}{3}~\cdot~ \PP[U] ~~=~~ \frac{1}{3}       \]
\end{proof}

We will show that the number of bad epochs at a level $i$ could exceed the number of good
epochs at level $i$ by at most $O(\log n)$ with very high probability.
Notice that the number of epochs created at level $i$ is itself a random variable.
During any update, the number of epochs that will be created at level $i$ depends upon the past updates in the graph and the random bits picked during their processing. However, Lemma \ref{lemma:generic-bad-epoch-prob} implies that each newly created epoch at level $i$ will be bad with probability at most $1/3$ independent of these events.
%Also note that the number of epochs created at level $i$ during any $t$ updates is trivially $O(nt)$.
Hence, the sequence of epochs at level $i$ can be seen as an instance of the asymmetric random walk
as established in the analysis of the 2-level algorithm.
%The walk starts at location
%$3\log_3 n$ to the right of the origin. Each step of the walk is one unit to the right of the current location with probability $3/4$ or one unit to the left with probability $1/4$ independent of the past moves. We need to find the probability that the walk ever reaches the origin in at most $nt$ steps. It follows from Lemma \ref{lemma:asymmetric-random-walk} that the probability of this event is less than $1/n^3$.
So the bad epochs at any level $i$ may exceed the good epochs by $2 \log_2 n$ with probability at most $1/n^2$. There are $O(\log n)$ levels in the hierarchy. Hence we get the following lemma using {\em union bound}.
%therefore, the bad epoch may exceed good epochs $b\log n$ at any level with probability at most $(\log n)/n^3<1/n^2$. Hence we can state the following lemma.
\begin{lemma}
For every level $i\le \textsc{l}_0$, the number of bad epochs will not exceed the number of good epochs by more than $2 \log_2 n$ with probability at least $1-(\log n)/n^2 > 1-1/n$.
\label{lemma:good-and-bad-epochs-generic}
\end{lemma}

\begin{center}
\begin{figure}
\psfrag{a}{$a$}
\psfrag{b}{$b$}
\psfrag{c}{$c$}
\psfrag{d}{$d$}
\psfrag{g}{$g$}
\psfrag{good}{A natural good epoch}
\psfrag{bad}{A bad epoch}
\psfrag{induced}{An induced epoch}
\psfrag{(i)}{(i)}
\psfrag{(ii)}{(ii)}
\psfrag{i}{$i+1$}
\psfrag{i-1}{$i$}
\centerline{\epsfysize=100pt \epsfbox{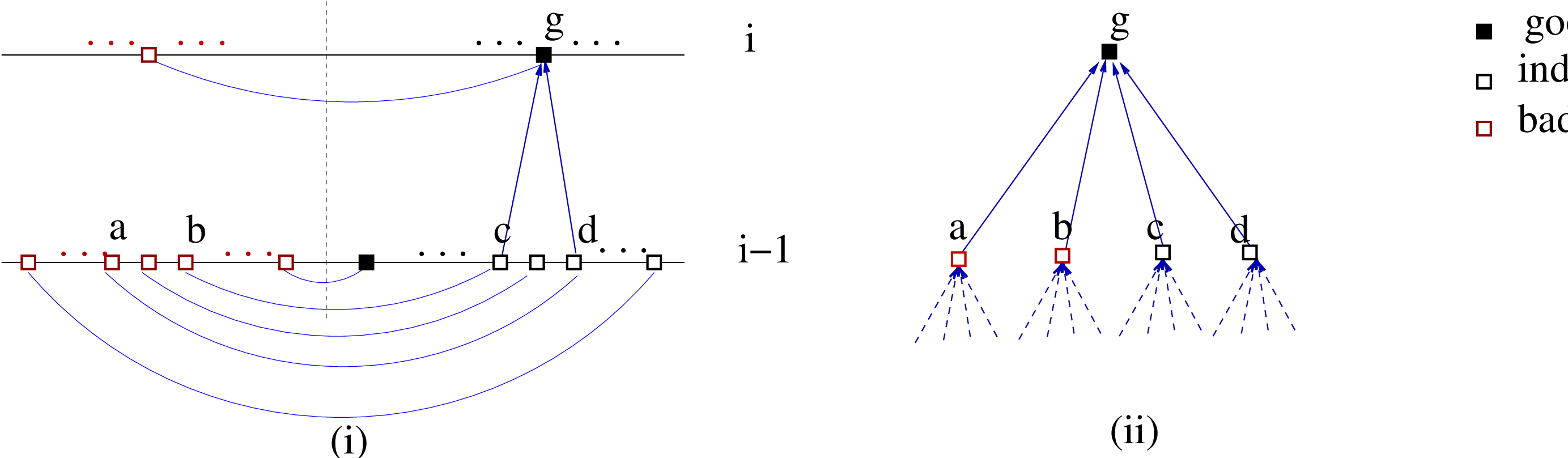}}
\caption{(i) Mapping between bad and good epochs at level $i$ ~(ii) Assigning at most 4 epochs from lower levels to an epoch.}
\label{figure:level-i-generic}
\end{figure}
\end{center}

Let us temporarily exclude the maximum surplus of $O(\log n)$ bad epochs at
each level from our analysis.
%Consider the epochs at a level $i$.
Consequently, it follows from Lemma \ref{lemma:good-and-bad-epochs-generic} that each bad epoch at a level
can be mapped to a good epoch at the same level in a unique manner -
see Figure \ref{figure:level-i-generic}(i). Also the creation of each epoch at a level $i+1$ can terminate at most two (induced) epochs at lower
levels as stated in Lemma \ref{lemma:atmost-2-induced-epochs}. Using this fact and the mapping between the good and bad epochs at a level, we can construct a forest whose nodes will be the epochs terminated across all levels during the algorithm.
% (excluding the surplus of $O(\log n)$ bad epochs at each level).
%This forest will provide a neat analysis for the total computation time of the generic algorithm spent in processing $t$ updates.
The intuition for defining this forest is that eventually the computation
cost of a bad epoch or an induced epoch will be charged to a good natural epoch. Since a good natural epoch has sufficiently {\em large} number of edge
deletions associated with it, these edge deletions can be charged to pay
for all the computation carried out by our algorithm.

With this intuition, we now provide the construction of the forest by defining parent of each epoch using the following rules.\\

\noindent
\begin{enumerate}
\item
Parent of each induced epoch is the epoch at the higher level whose creation led to its termination.
\item
Parent of a good epoch is itself (hence it is the root of its tree).
\item If a bad epoch is mapped to an induced epoch, then its parent is the same as the parent of the induced epoch.
Otherwise, it is the parent of itself (hence it is the root of its tree).
\end{enumerate}

It follows from rule 1 and 3 (the if part) that with an epoch at a level,
at most 4 epochs from lower levels can be associated. Hence each node in the
forest will have at most four children. See Figure \ref{figure:level-i-generic}(ii). Moreover, the root of each tree in the forest of epochs is either a bad epoch or a good natural epoch.
%We analyse the total update time of the algorithm by analysing the computation associated with each such tree.
Using Lemma \ref{lemma46}, the computation cost $C(i)$ associated with a tree of epochs whose root is at level $i$ obeys the following recurrence for some constant $a$.
\[ C(i) = a 4^i + 4C(i-1)\]
The solution of this recurrence is $C(i)=O(i4^i)$.
It follows from Lemma \ref{lemma:good-and-bad-epochs-generic} and rule 3(Otherwise part) that the trees rooted at good natural epochs at a level $i$ are at least the number of trees rooted at bad epochs at level $i$. Hence, it suffices to analyze the computation cost associated with all the tree rooted at good natural epochs. Now for each good natural epoch at a level $i$, there are at least $4^{i}/3$ edge deletions associated uniquely to it.
This natural epoch will be charged for the computation cost $C(i)=O(i4^i)$ associated with the tree
rooted at it.
So if $t$ is the total number of updates in the graph, then the computation cost associated with all epochs in the forest is $O(t \log n)$.
%All that is left is to analyse the computation cost associated with the surplus of $O(\log n)$ bad epochs at each level.
The computation cost associated with surplus bad epochs at all levels is
 $O(\sum_i i4^i \log n) = O(n\log^2 n)$. Hence with high
probability the computation cost for processing $t$ edge updates by the algorithm is $O(t\log n + n \log^2 n)$.
%\begin{theorem}
%For any sequence of $t$ updates in a graph on $n$ vertices, the total time taken for maintaining a maximal matching by the  algorithm
%is bounded by $O(t \log n + n\log^2 n)$ with high probability.
%\label{theorem:final-theorem-generic-high-probability}
%\end{theorem}
%
%2016-begin
%Let us now analyze the computation {\em charged} to all those epochs which are
%alive at the end of $t$ updates. Consider an epoch of a matched edge, say
%$(u,v)$ at level $i$, such that $(u,v)$ is still in the matching after $t$
%updates.
%
%By Lemma~\ref{lemma47}, computation charged to
%epoch of $(u,v)$ is $O(i2^i)$. If $\OO_v^{init}$ were the edges owned by $v$
%at the start of the epoch, then $|\OO_v^{init}| \ge 2^i$ at the start of the
%epoch. We can say that the computation charged is
%$O( i |\OO_v^{init}|) \le \ O( (\log n) t_v)$
%where $t_v$ is the total number edge updates on $v$. Since a vertex can
%be part of only one live epoch, the total computation cost charged to
%all the live epochs is $\sum_v O(t_v \log n) =   O(t \log n)$.
%2016-end
%
This also implies %from Theorem \ref{theorem:final-theorem-generic-high-probability}
that the total expected update time is $O(t\log n)$ for
$t=\Omega(n\log n)$. In the following subsection, we will
establish $O(t\log n)$ bound on the expected update time for all values of $t$.
%---------------------------------------------------------------------------------------------------%
                \subsubsection{Expected value of the total update time}
%---------------------------------------------------------------------------------------------------%

During a sequence of $t$ updates in the graph, various epochs get created by various vertices at various levels.
Let $X_{v,i,k}$ be a random variable which is 1 if $v$ creates an epoch at level $i$ at update step $k$, otherwise it is 0. We denote this
epoch as \textsc{epoch}($v,i,k$). Let $\OO_v^{init}$ denote the edges that $v$ owned at the time of the creation of the epoch.
%So, this random variable depends only the random coin tosses till update step $t$.
Let $Z_{v,i,k}$ denote the number of edges from $\OO_v^{init}$ that are deleted during the epoch. (If \textsc{epoch}($v,i,k$) is not created,
$Z_{v,i,k}$ is defined as 0). The key role in bounding the expected running time is played by a random variable $B_{v,i,k}$ defined as follows:

\begin{equation}
    B_{v,i,k}=
    \begin{cases}
      (8Z_{v,i,k} - 2\cdot 4^i)X_{v,i,k}~ & \text{if}~\textsc{epoch}(v,i,k) ~\text{is natural} \\
      (4^{i+1} - 2 \cdot 4^i)X_{v,i,k}~ & \text{if}~\textsc{epoch}(v,i,k)~\text{is induced} \\
    \end{cases}
  \end{equation}

First observe that $B_{v,i,k}=0$ if $X_{v,i,k}=0$. Else (if $X_{v,i,k} = 1$),
the random variable $B_{v,i,k}$ can be seen as credits associated with \textsc{epoch}($v,i,k$) to be used for paying its computation cost.
For a natural epoch, the credits is defined in terms of the edges deleted during the epoch.
So we define $B_{v,i,k}$ to be $8Z_{v,i,k}$. However, we need to discount for the two epochs at lower levels that may get terminated due to \textsc{epoch}$(v,i,k)$. To this end, from the term, we deduct $2 \cdot 4^i$. Similarly, if \textsc{epoch}$(v,i,k)$ is an induced epoch, then it gets $4^{i+1}$ credits from the epoch that destroyed it. But here again
%By Observation \ref{obs:obs1}, even induced  epoch may destroy two epochs at lower level, so again
we need to discount for the two epochs at lower levels that might be terminated by it. To this end, we again deduct $2\cdot 4^i$ from
$4^{i+1}$. The following lemma gives a bound on $\sum B_{v,i,k}$.

­\begin{lemma}
\label{lem:lem-main}
 $\sum_{v,i,k} B_{v,i,k} \le 8 \sum_{v,i,k} Z_{v,i,k} \le 8t$, where $t$ is the total number of updates in the graph.
\end{lemma}
\begin{proof}
% Consider the execution of our algorithm using fixed set of coin tosses  from the sample space, say $\sigma$.
We need to analyze the sum of $B_{v,i,k}$'s for all those $(i,v,k)$ values for which the \textsc{epoch}$(i,v,k)$ got
created.
%\complain{Should we point out explicitly that there is something
%special about the highest level where there cannot be an induced epoch. [Manoj] Not needed in my view}
If this epoch is an induced epoch, it can be associated with an epoch, say \textsc{epoch}$(v',i',k')$, at a higher level $i' > i$ whose
creation destroyed it. %(Note that $X_{v,i,k} = X_{v',i',k'} =1)$.
Notice that the negative $4^{i'}$ term in $B_{v',i',k'}$ cancels out the positive $4^{i+1}$ term in $B_{v,i,t}$. Hence, the contribution of induced epochs in $\sum_{v,i,k}B_{v,i,k}$ is nullified and all that remains is the sum of terms $8Z_{v,i,k}$ for each natural epoch.
Hence $\sum_{v,i,k}B_{v,i,k} \le 8 \sum_{v,i,k} Z_{v,i,k} $. An edge deletion is associated with an epoch in a unique manner, so will contribute to exactly one $Z_{v,i,k}$. Therefore,
$\sum_{v,i,k} Z_{v,i,k}$ is upper bounded by the total number of edges deleted.
%When an edge is deleted it can be a part of at most 2 epochs.
%Hence $ \sum_{v,i,k} Z_{v,i,k} \le t$. This completes the proof.
\end{proof}

\begin{corollary}
\label{cor:cor1}
   $\sum_{v,i,k} \EX[B_{v,i,k}] \le 8 t$
\end{corollary}

\begin{lemma}
\label{lem:lem1}
 For all $i,v, k$, \ $\EX[B_{v,i,k}] \ge \PP[X_{v,i,k}=1]\cdot 4^i$.
\end{lemma}
\begin{proof}
Since $X_{v,i,k}$ is an indicator random variable,
$\EX[B_{v,i,k} ] = \PP[X_{v,i,k}=1] ~\EX[B_{v,i,k} |\ X_{v,i,k}=1] $.
We will first estimate
$\EX[B_{v,i,k} |\ X_{v,i,k}=1]$, that is, the expected value of $B_{v,i,k}$ given that \textsc{epoch}($v,i,k)$ got created.
%We accomplish this task using Lemma \ref{lemma:key-lemma-generic-algo} as follows.

Let $({\cal U},P)$ be the probability space of all the update sequences associated with this epoch and let $U\in {\cal U}$ be any
update sequence.
Suppose among the updates in $U$ that precede the update associated with $v$, only $d$ are edge deletions. It follows from Lemma \ref{lemma:key-lemma-generic-algo} that the matched edge of $v$ is distributed uniformly over $\OO_v^{init}$. So \textsc{epoch}($v,i,k)$  will be an induced epoch with probability $(|\OO_v^{init}| - d)/|\OO_v^{init}|$
and in that case $B(v,i,k)$ will be $4^{i+1}-2\cdot 4^i$. If the epoch is
natural, it could be due to any one of the $d$ edge deletions present in $U$. In that case the expected value of $B_{v,i,k}$ will be $1/d \sum_{j=1}^d (8j -2\cdot 4^i)\ge 4d - 2\cdot 4^i$. Considering the cases of induced and natural epoch together,
%the conditional expectation of
\begin{eqnarray*}
\EX[B_{v,i,k}~|~U]  =
\frac{|\OO_v^{init}|-d}{|\OO_v^{init}|}(4^{i+1}-2\cdot 4^i) ~+~\frac{d}{|\OO_v^{init}|}(4d - 2\cdot 4^i)
               &=& 2\cdot 4^i - \frac{4^{i+1}d-4d^2}{|\OO_v^{init}|} \\
               & \ge &  2\cdot 4^i - \frac{4^{i}\cdot 4^i}{|\OO_v^{init}|}   ~\mbox{(for all values of $d$)}
\end{eqnarray*}
Therefore
\[ \EX[B_{v,i,k}]~
    =~ \sum_{U\in {\cal U}} \EX[B_{v,i,k}~|~U]~\cdot~ \PP[U] ~
    \ge~ \left( 2\cdot 4^i - \frac{4^{i}\cdot 4^i}{|\OO_v^{init}|}  \right) \cdot \sum_{U\in {\cal U}} \PP[U] ~
    =~ 2\cdot 4^i - \frac{4^{i}\cdot 4^i}{|\OO_v^{init}|}
\]
Since $|\OO_v^{init}| \geq 4^i$, for level $i$,
the result follows.
\end{proof}

Let $W_{v,i,k}$ be a random variable that corresponds to the value of the
computation cost of \textsc{epoch}($v,i,k$) if the epoch
is created and is 0 otherwise.
Notice that the computation cost of an epoch at level $i$ is $c4^{i+1}$
for some constant $c$.
So, $\EX [W_{v,i,k}] = \PP[X_{v,i,k} = 1] c 4^{i+1}$. Therefore, using Lemma \ref{lem:lem1},
\begin{equation}
\EX[W_{v,i,k}] \le 4c \EX[B_{v,i,k}]
\end{equation}
Using the above equation and Corollary \ref{cor:cor1}, the total expected computation cost  associated with all epochs that get
destroyed during the algorithm  can be bounded by $O(t)$ as follows.
\[ \sum_{v,i,k} \EX[W_{v,i,k}] \le \sum_{v,i,k} 4 c\EX[B_{v,i,k}] \le 32 ct  = O(t)\]
Since for each update in the graph, we incur $O(\log n)$ time to update $\phi$ at various levels, there is an $O(t \log n)$ overhead
for $t$ updates. We can thus conclude with the following theorem.
%\begin{theorem}
%For any sequence of $t$ updates in a graph on $n$ vertices, the expected time taken for maintaining a maximal matching by the algorithm is $O(t \log n)$.
%\label{theorem:final-theorem-generic-expected}
%\end{theorem}

\begin{theorem}
Starting with a graph on $n$ vertices and no edges, we can maintain a maximal
matching for any sequence of $t$ updates in $O(t\log n)$ time in expectation
and $O(t \log n + n \log^2 n)$ with high probability.
\label{theorem:final-theorem-generic-algo}
\end{theorem}

%We shall now state our insight for the generic algorithm. This insight played the key role in extending
%Lemma \ref{lemma:key-lemma-level-2-algorithm} in the 2-level algorithm to  Lemma
%\ref{lemma:key-lemma-generic-algo} in the generic algorithm.

%---------------------------------------------------------------------------------------------------------------%
             \subsection{Proof of Lemma \ref{lemma:key-lemma-generic-algo}}
%---------------------------------------------------------------------------------------------------------------%
\label{section:proof-of-key-lemma-generic-algo}
Our algorithm use randomization to maintain maximal matching. After any given sequence of updates, there is a set of possible maximal matchings that the algorithm may be maintaining and there is probability distribution associated with these
maximal matchings. So it is useful to think about the probability space of these matchings as the algorithm proceeds while processing a sequence of updates.

We introduce some notations first. For any matching ${\cal M}$ maintained at any stage by our algorithm, let ${\cal M}_i$ denote the matching at level $i$. Let ${\cal M}_{>i}=\cup_{j>i}{\cal M}_j$ denote the matchings at all levels $>i$.
Let $V_i$ denote the set of all the vertices belonging to levels in the range $\in[-1,i]$. We now extend the notations to incorporate the updates in the graph. For any $k\ge 1$, let $G(k)$ denote the graph after a given sequence of $k$ updates and
let ${\cal M}(k)$ denote the maximal matching of $G(k)$ as maintained by our
algorithm. Let ${\cal M}_{>i}(k)$ denote the matching at all levels $>i$ after a given sequence of $k$ updates.

After processing certain number of updates by the algorithm,
suppose $M$ and $M'$ are any two matchings possible such that
$M_{>i}=M_{>i}'$.  Consider any single update in the graph at this stage.
In order to process it, suppose we carry out two executions $I$ and $I'$ of our algorithm with the initial matching being $M$ and $M'$ respectively.
That is, ${\cal M}(0)=M$ in the execution $I$ and ${\cal M}(0)=M'$ in the execution $I'$.
Our claim is that the probability distribution of matching at levels $>i$ will be identical at the end of both the executions.
More precisely, for any maximal matching $\mu(1)$ on a subset of vertices in graph $G(1)$,
\[ \PP[{\cal M}_{>i}(1) = \mu(1) | {\cal M}(0)=M] ~~=~~ \PP[{\cal M}_{>i}(1) = \mu(1) | {\cal M}(0)=M']\]
\noindent
In order to establish our claim, we shall crucially exploit the following lemma.
\begin{lemma}
For both the matchings $M$ and $M'$, $\phi_{v}(j)$ is the same for each $v\in V$ and $j>i$.
\label{lemma:initial-M'-and-M''}
\end{lemma}
\begin{proof}
It is given that $M_{>i}=M_{>i}'$. This implies that for each level $j>i$ the sets of vertices present are identical in $M$ and $M'$.
Hence the set $V_i$ of all the vertices present at levels $\in [-1,i]$ is
identical in $M$ and $M'$. Hence for any vertex $v$, and any level $j>i$,
the set of
all the neighbours of $v$ at levels $<j$ is identical; notice that
$\phi_{v}(j)$ is just the cardinality of this set.
So it follows that $\phi_{v}(j)$ is the same for each vertex $v$ and each $j>i$.
\end{proof}

We shall now establish our claim for the deletion of an edge $e=(u,v)$. Establishing the claim for the insertion of an edge is similar.
Notice that our algorithm does not alter the matching if $e$ is not a matched edge.
If $e$ is a matched edge, a wave of free vertices originates from $\LL(e)$ and propagates downward.
The following fact follows from our analysis in Section \ref{procedure-process-free-vertices}.
\begin{enumerate}
\item[$F1$.] The algorithm won't alter the matching at level $> \LL(e)$ while processing the deletion of $e$.
\item[$F2$.] The matching is updated in the decreasing order of levels, and once the updating of the matching at a level is complete,
the matching at that level will remain unchanged during the updates of the matching at lower levels.
\end{enumerate}

It follows from the description of $M$ and $M'$ that either $\LL(e)$ is less than or equal to $i$
in both the matchings or $\LL(e)$ is the same in $M$ and $M'$.
Let us first consider the (easier) case when $\LL(e)\le i$ in $M$ as well as $M'$.
It follows from Fact $F1$ stated above that the only changes in matching $M$ and $M'$ will be at levels $\le i$. Hence the matching
${\cal M}_{>i}(1)$ will be identical at the end of both the executions $I$ and $I'$.
Let us now consider the more interesting case of $\LL(e)>i$.
Both the executions $I$ and $I'$ invoke the procedure \textsc{process-free-vertices}($\langle(u,\LL(e)), (v,\LL(e)) \rangle$) in this case.
%Both the executions begin with a queue at level= $\LL(e)$ containing the free vertices $\{u,v\}$ and invoke procedure
%\textsc{process-free-vertices}.
The reader is recommended to revisit this procedure from Section \ref{procedure-process-free-vertices} before proceeding further.

%Refer to the text preceding Remark \ref{remark:similarity-with-2-levels}
%
%This procedure iteratively does the following.
%\begin{enumerate}
%\item
%Pick the highest level free vertex, say $w$. Let $\ell$ be its level.
%\item
%$w$ gives up the ownership of its edges that are incident from level $\ell$ and is processed as follows.
%\begin{enumerate}
%\item
%If $w$ still owns at least $4^\ell$ edges, it picks a random edge from its owned edge and gets matched.
%\item
%If $w$ owns less than $4^\ell$ edges, it drops its level by 1.
%As a result, for each neighbour $z$ of $w$ from lower levels, $\phi_{\ell}(w)$ increases by one. Each such neighbour $z$ will rise to level
%$\ell$ if and only if $\phi_{\ell}(z)$ has become $4^\ell$.
%Each such rising vertex selects a random edge from its owned edges and gets matched.
%\end{enumerate}
%\end{enumerate}
%----------------Mention the following sentence just after the Procedure \textsc{process-free-vertices} in the SICOMP paper---------------------
%The reader may notice a stark similarity between the above procedure and the procedure used in the 2-level algorithm. In fact, a single
%iteration of the above procedure is exactly the same as the procedure that the 2-level algorithm executes to process a free vertex at level 1.
%-----------------------------------------------------------------------------------------------------------
In order to establish our claim about $I$ and $I'$, we shall establish the following. While the matching at levels $>i$ is being updated, for each step executed in $I$, the identical step can be executed in $I'$. Moreover, if the step in $I$ is executed with some probability,
the step will be executed with the same probability in $I'$ as well.
In order to show this, let us analyse the first iteration of the procedure \textsc{process-free-vertices}.
Both $I$ and $I'$ will process $u$ first. After disowning its edges from its present level, $u$ owns the same set of edges in both the executions. Thereafter, $u$ will either stay at the same level or fall by one level. If $u$ stays at the same level, it
chooses a random edge to get matched. The probability that any specific random edge is picked by $u$ is the same in both the executions.
Let us consider the case that $u$ falls by one level. For each neighbour $z$ of $u$, it follows from Lemma \ref{lemma:initial-M'-and-M''} that $\phi_\ell(z)$ is the same in the case of $M$ and $M'$. Hence the set of vertices rising to level $\ell$ are the same in both the executions.
In addition, the set of edges that each such vertex owns on rising to level $j$ is also the same, hence, the probability that any specific random mate is picked is the same in both the executions. So each update in $M$ and $M'$ is equally likely during the processing of $u$. The reader may note that after each such identical update in $M$ and $M'$, the matchings are identical
at each level $>i$. Hence, Lemma \ref{lemma:initial-M'-and-M''} holds again for the updated matchings.
%Therefore, the above analysis gets carried over to the second and every subsequent update step carried out by $I$.

Unlike the first iteration, a generic iteration of the procedure \textsc{process-free-vertices} may have free vertices at levels
$\le\textsc{level}(e)$ that are kept in respective queues at these levels.
Suppose in the beginning of any such iteration of the procedure \textsc{process-free-vertices} there are two possible configurations
such that the matching as well as the queue storing the free vertices are identical at each level $>i$ but differ at levels $\le i$.
Lemma \ref{lemma:initial-M'-and-M''} will hold for these configurations as well.
Therefore, along {\underline{exactly}} the same lines as the first iteration analysed above, it can be shown that every update in the
matching at level $>i$ will be carried out with the same probability during any generic iteration for any two configurations that match
at all levels $>i$.

Therefore, each sequence of updates in the matching is equally likely in both the executions $I$ and $I'$ till the last free vertex at level $i+1$ is processed. Henceforth, the two executions may differ. But as follows from Fact $F2$, it will affect only the matching at levels
$\le i$ and there won't be any change in the matching at higher levels.

%So we can conclude the following. Let $M$ and $M'$ be any two valid initial matchings that are identical at each level $>i$. After processing any single update in the graph by the generic algorithm starting with $M$ and $M'$ respectively, the probability distributions of the updated matchings at each level $>i$ will also be identical.
This concludes our claim for a single update. This claim can be invoked
appropriately for a sequence of updates giving us the following theorem.
\begin{theorem}
Let $M$ and $M'$ be any two matchings possible by our algorithm at any time such that $M_{>i}=M_{>i}'$.
For any sequence of $t$ update in the graph, suppose we carry out two executions $I$ and $I'$ of our algorithm with the initial matching being $M$ and $M'$ respectively. The probability distribution of matching at every level $>i$ will be identical at the end of both the executions. That is,
\[ \PP[{\cal M}_{>i}(t) = \mu(t),\ldots, {\cal M}_{>i}(1)=\mu(1) | {\cal M}(0)= M]~~ =~~
\PP[{\cal M}_{>i}(t) = \mu(t),\ldots, {\cal M}_{>i}(1)=\mu(1) | {\cal M}(0)=M']\]
where $\mu(j)$, for $1\le j \le t$, is any maximal matching on a subset of vertices in the graph $G(j)$.
\label{theorem:identical-distribution-of-matchings-beyond-level-i}
\end{theorem}

\noindent
For the proof of Theorem \ref{theorem:identical-distribution-of-matchings-beyond-level-i}, we shall apply the argument for single update inductively and use the following lemma from elementary probability theory.
\begin{lemma}
Suppose $A,B,C$ are three events defined over a probability space $(\Omega,P)$. Then,
\[ \PP[A\cap B ~|~C] = \PP[A~|~B \cap C] \cdot \PP[B~|~C] \]
\label{lemma:chain_cond}
\end{lemma}

Let us define events $C$ as ${\cal M}(0)= M$ and $C'$ as ${\cal M}(0)= M'$.
We have shown that $\PP [ {\cal M}_{>i}(1)=\mu(1)~ |~ C ] =\PP[{\cal M}_{>i}(1)=\mu(1) ~|~ C' ]$.
If we define event $B$ as ${\cal M}_{>i}(1)=\mu(1)$ then by another application of the arguments that we used for a single update,
\[ \PP [ {\cal M}_{>i}(2)=\mu(2)~ |~ B , C ] = \PP [ {\cal M}_{>i}(2)=\mu(2) ~|~ B , C' ]  \]
Applying Lemma \ref{lemma:chain_cond}, we get
\[\PP [ {\cal M}_{>i}(2)=\mu(2) , B~ |~ C]=  \PP [ {\cal M}_{>i}(2)=\mu(2) ~|~ B , C]\cdot \PP[ B~ |~ C]  \]
Since $\PP [ B~|~ C ] = \PP [ B | C']$, it follows that
\[ \PP [ {\cal M}_{>i}(2)=\mu(2), B ~|~ C] = \PP [ {\cal M}_{>i}(2)=\mu(2), B~ |~ C' ]  \]
The above argument can be inductively applied for every subsequent update.
This completes the proof of Theorem \ref{theorem:identical-distribution-of-matchings-beyond-level-i}
\subsubsection{Connection to the analysis}
We first state two lemmas from elementary probability theory that deal
with the independence of events. For the sake of completeness, the proof of these lemmas is given in Appendix.
%
%\subsection{Two lemmas from probability theory}
%We now state two lemmas from elementary probability theory.

\begin{center}
\fbox{\parbox{6.5in}{
The first lemma deals with conditional probability.
%The first lemma states that if event $A$ is independent of a set
%of mutually exclusive events, then $A$ is independent of the union of the events of the set as well.
%
\begin{lemma}
Let $A$ be an event and $B_1,\ldots,B_k$ be $k$ mutually exclusive events defined over a probability space
$(\Omega,P)$. If $\PP[A~|~B_j]=\rho$ for each $1\le j \le k$, then $\PP[A~|~C]=\rho$ where event $C= \cup_{j} B_{j}$.
\label{lemma:conditional-probability}
\end{lemma}

The second lemma deals with independence of events.
Let $A$ and $B$ be two events defined over a probability space $(\Omega,P)$.
$A$ is said to be independent of $B$ if $\PP[A~|~B] = \PP[A~|~\bar{B}] = \PP[A]$. Alternatively, $\PP[A\cap B] = \PP[A]\cdot \PP[B]$.
The notion of independence gets carried over from events to random variables in a natural manner as follows.
\begin{definition}
An event $A$ is said to be independent of a random variable $X$ if for each $x\in X$, $\PP[A~|~ X=x]= \PP[A]$.
\label{def:independence}
\end{definition}

\begin{lemma}
Suppose $A$ is an event and $X$ be a random variable defined over probability space $(\Omega,P)$. If $A$ is independent of $X$, then
for each $x\in X$,\\
\hspace*{2.5in}$ \PP[X=x~|~A] = \PP[X=x] $
\label{lemma:independence-is-symmetric-lemma}
\end{lemma}
}}
\end{center}
Now we shall establish the connection of Theorem \ref{theorem:identical-distribution-of-matchings-beyond-level-i} to the anlysis
of our algorithm. In particular, we shall use this theorem to prove Lemma \ref{lemma:key-lemma-generic-algo}.
Suppose a vertex $v$ creates an epoch at level $i$ while the algorithm processes $k$th update in the graph for any $k< t$.
%To analyze the duration of any epoch, we would like to
%compute the probability of this epoch being good or bad, natural or induced.
%For this purpose,
We shall analyse the probability space of the future matchings starting from the time just before the creation of this epoch.

While creating its epoch, $v$ chooses its mate randomly uniformly out of
$\OO_v^{init}$. Clearly, the change in the matching at levels $\le i$ will depend on the mate that $v$ picks.
Let ${\mathbf{M}}$ be the set of all possible matchings once the algorithm completes the processing of the $k$th update.
Now notice that all matchings from the set ${\mathbf{M}}$ are identical at each level $>i$.
So it follows from Theorem \ref{theorem:identical-distribution-of-matchings-beyond-level-i} that for any two matchings $M,M'\in {\mathbf{M}}$,
\begin{tabbing}
    $\PP[{\cal M}_{>i}(t) = \mu(t),\ldots,$\=$ {\cal M}_{>i}(k+1)=\mu(k+1)~ |~ {\cal M}(k)=M]$\\
    \>$ = \PP[{\cal M}_{>i}(t) = \mu(t),\ldots, {\cal M}_{>i}(k+1)=\mu(k+1) ~| ~{\cal M}(k)=M']$
\end{tabbing}
Let this conditional probability be $\rho$.
For each $(v,w)\in \OO_v^{init}$, there may be several matchings in ${\mathbf{M}}$ in
which $v$ is matched to $w$.
By applying Lemma \ref{lemma:conditional-probability}, the following equation holds for every $(v,w)\in \OO_v^{init}$,
\[
    \PP[{\cal M}_{>i}(t) = \mu(t),\ldots, {\cal M}_{>i}(k+1)=\mu(k+1) ~|~\textsc{mate}(v)=w] ~=~ \rho
   % \PP[{\cal M}_{>i}(t) = \mu(t),\ldots, {\cal M}_{>i}(k+1)=\mu(k+1) ~|~ \textsc{mate}(v)=w']
\]
Since this probability is the same for each $(v,w)\in \OO_v^{init}$, so using Definition \ref{def:independence}, it follows that
the matchings at levels $>i$ during any sequence of updates is independent of the mate that $v$ picked during the creation of its epoch. Now applying Lemma \ref{lemma:independence-is-symmetric-lemma} we get the following lemma.
\begin{lemma}
Suppose a vertex $v$ creates an epoch at level $i$ while the algorithm processes $k$th update in the graph. Consider any sequence of updates in the graph. The mate picked by $v$ while creating the epoch is independent of the sequence of matchings at levels $>i$ computed by the algorithm while processing these updates. That is, for any $t>k$, and any $(v,w)\in \OO_v^{init}$,
\[
       \PP[\textsc{mate}(v)=w | {\cal M}_{>i}(t) = \mu(t),\ldots, {\cal M}_{>i}(k+1)=\mu(k+1)] ~~
       =~~\PP[\textsc{mate}(v)=w]~~=~~ \frac{1}{|\OO_v^{init}|}
\]
\label{lemma:prob-distr-independent-of-mate-of-v}
\end{lemma}

Consider any given sequence of $t$ updates in the graph.
Subsequent to the time $v$ creates an epoch at level $i$ during $k$th update, let
${\mathbf{\mu}} = \langle \mu(k+1),\ldots,\mu(t) \rangle$ be the sequence of matching at levels $>i$ as computed by the algorithm. Notice that the upward movement of $v$ and each $(v,w)\in \OO_v^{init}$ after the creation of epoch is captured precisely by the corresponding update in the matching at level $>i$. Therefore, using ${\mathbf{\mu}}$ we can define the update sequence associated with the epoch as follows.
Consider an edge $(v,w)\in \OO_v^{init}$ and let $\ell$th update in the graph be the deletion of $(v,w)$.
Let $j<\ell$ be the smallest integer such that $w\in \mu_j$, that is, $w$ appears in the
matching at level $>i$ while processing of $j$th update in the graph, then the update associated with $(v,w)$ is the upward movement. If no such $j$ exists, the update associated with $(v,w)$ is its deletion. Likewise, we define the update associated with $v$. The update sequence $U$ for the epoch is the sequence of these updates on $v$ and the edges of $\OO_v^{init}$ arranged in the chronological order.

For an update sequence $U$ associated with an epoch, there may exist many sequences $\{\mu_1,\ldots,\mu_q\}$ such that for each of them, the update sequence associated with the epoch is $U$. It follows from Lemma \ref{lemma:prob-distr-independent-of-mate-of-v} that the mate picked by
$v$ during its epoch is independent of each such sequence ${\mu}_r, 1\le r\le q$.
Therefore, using Lemma \ref{lemma:conditional-probability},  the mate picked by $v$ during its epoch is independent of $U$ as well.
Thus we have established the validity of Lemma \ref{lemma:key-lemma-generic-algo}.

\section{A tight example}
\begin{figure}[h]
\psfrag{n/2}{$n/2$}
\centerline{\epsfysize=140pt \epsfbox{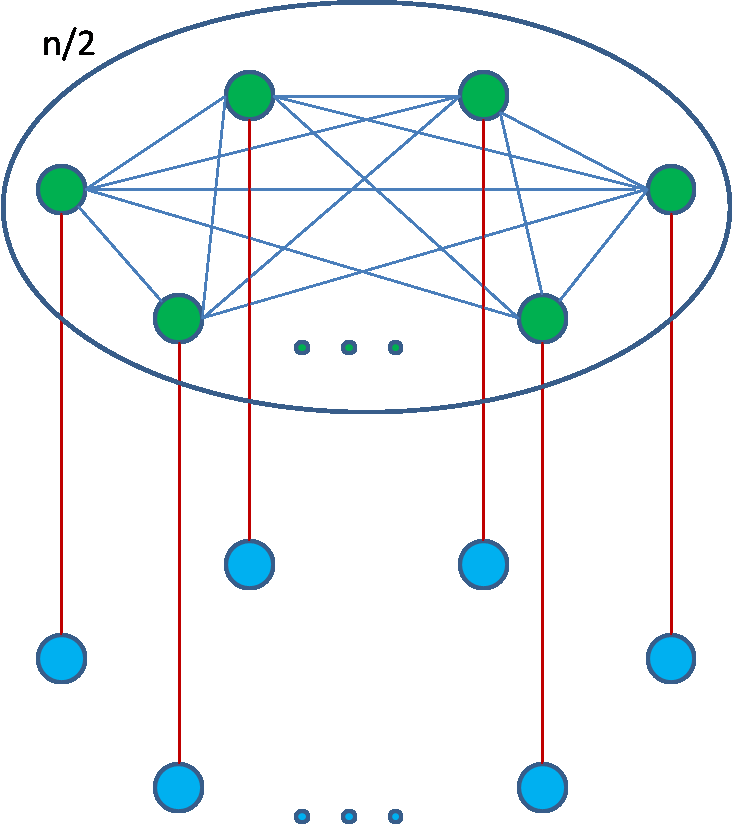}}
\caption{An example where our algorithm gives a 2-approximation. The vertices
on top are in $V$ and form a complete graph. The vertices at the bottom of the figure are in $W$.}
\label{Figure8}
\end{figure}

We tested our algorithm on random graphs of various densities and found that
the matching maintained is very close to the maximum matching. This
suggests that our algorithm might be able to maintain {\em nearly} maximum
matching for dynamic graphs appearing in various practical applications.
However, it is not hard to come up with an update sequence such that at the
end of the sequence, the matching obtained by our algorithm is strictly
half the size of maximum matching. In other words, the approximation
factor 2 for the matching maintained by our algorithm is indeed tight.
We present one such example as follows (see Figure \ref{Figure8}).

Let $G( V \cup W  , E)$ be a graph such that $V = \{v_1, v_2, \dots , v_n\}$
and $W = \{w_1, w_2, \dots, w_n\}$ for some even number $n$. Consider the
following update sequence. In the first phase, add edges between every pair of
vertices present in $V$.
This results in a complete subgraph on vertices of $V$.
The size of any maximal matching on a complete graph of size $n$ is $n/2$.
After the first phase of updates ends, the size of matching obtained by our
algorithm is $n/2$. In the second phase, add edge $(v_i,w_i)$ for all $i$.
Note that the degree of each $w_i$ is one at the end of the updates.
Let us now find the matching which our algorithm maintains.
Let $(v_i,v_j)$ be an edge in the matching after phase 1. Note that both
these endpoints are at a level greater than $-1$. A vertex in $W$ is at level $-1$
as it does not have any adjacent edges after phase 1. When an edge
$(w_i,v_i)$ is added, since $v_i$ is at a higher level than $w_i$,
$v_i$ becomes the owner of this edge. The second invariant of $v_i$ is not
violated after this edge insertion and nothing happens at this update step
and $w_i$ still remains at level $-1$. Using same reasoning, we can show that
$w_j$ also remains at level $-1$ after the addition of edge $(v_j,w_j)$.
So matching maintained by the algorithm remains unchanged. It is easy to
observe that the maximum matching of the graph $G$ now has size $n$ which is
twice the size of the matching maintained by our algorithm.

\section{Postscript}
We presented a fully dynamic randomized algorithm for maximal matching which 
achieves expected amortized $O(\log n)$ time per edge insertion or deletion. 
An interesting question is to explore how crucial randomization is for dynamic maximal matching.

Subsequent to an earlier version of this paper \cite{DBLP:journals/siamcomp/BaswanaGS15}, 
Bhattacharya et al. \cite{DBLP:conf/stoc/BhattacharyaHN16} almost answered this question in affirmative by designing a determinstic algorithm that maintains $(2+\epsilon)$-approximate matching in amortized $O(poly (\log n, 1/\epsilon)$ update time. 
Another interesting question is to explore whether we can achieve $O(1)$ amortized update time. Very recently Solomon \cite{DBLP:journals/corr/Solomon16} answered this question in affirmative as well by designing a randomized algorithm that takes $O(t + n \log n)$ update time with high probability to process any sequence of $t$ edge deletions. Though the basic building blocks of his algorithm are the same as ours, 
the two algorithms are inherently different and so are their analysis.

In our algorithm, a vertex may rise to a higher level and create a new epoch even when its matched
edge is intact. But  the algorithm of Solomon \cite{DBLP:journals/corr/Solomon16} of takes a lazy approach to maintain the hierarchy of vertices
wherein a vertex is processed only when it becomes absolutely necessary. Another crucial difference is the following.
Our algorithm maintains a function $\phi_v(j)$ for each vertex $v$ and each level $j$.
This function is used to ensure an invariant that each vertex $v$ is at the highest possible level $\ell$ such that the edges incident from
lower levels is at least $4^{\ell}$. An important property guaranteed by this invariant is that the mate of a vertex while creating an epoch
is independent of the update sequence associated with the epoch. The analysis of our algorithm crucially exploits this property. 
However, the explicit maintenance of $\phi_v(j)$ imposes an overhead of $\Theta(\log n)$ in the update time. In order to achieve $O(1)$ update time, Solomon \cite{DBLP:journals/corr/Solomon16} gets rid of the maintenance of $\phi_v(j)$ by taking a lazy approach and a couple of new ideas. As a result, unfortunately, the property of our algorithm no longer holds for the algorithm of Solomon \cite{DBLP:journals/corr/Solomon16} -  indeed there is dependence
between the update sequence associated with an epoch created by a vertex and the random mate picked by it. 
Solomon \cite{DBLP:journals/corr/Solomon16} makes use of a new concept called {\em uninterrupted duration} of an epoch that bypasses the
need of our property for the analysis. His analysis can be adapted to our 
algorithm as well and can be viewed as the correct counterpart of Lemma
4.10 in \cite{DBLP:journals/siamcomp/BaswanaGS15}. 
However, our new analysis has its own merits
since it is based on an insightful property of our algorithm which we believe is of its own independent interest and importance.

Subsequent to the publication of the \cite{DBLP:journals/siamcomp/BaswanaGS15} 
there has been interesting progress in the area of dynamic 
matching with approximation less than 2 \cite{DBLP:conf/icalp/BernsteinS15, DBLP:conf/soda/BernsteinS16, NeimanS12, GuptaP13}, and
dynamic weighted matching \cite{AnandBGS12,AnandBGS-Arxiv12,GuptaP13}.

One of the technical challenges in theoretical computer science is to prove 
lower bounds for algorithmic problems. Recently there has been some 
progress on proving conditional lower bounds for dynamic graph algorithms
\cite{AbboudW14, DBLP:conf/stoc/HenzingerKNS15}. 
In the light of the lower bound presented by Abboud and Williams 
\cite{AbboudW14} based on $\Omega(n^2)$ hardness of the 3SUM problem,
it would be an interesting and challenging problem to see 
if $c$-approximate maximum matching for $c<2$ can be maintained in $o(n)$ 
update time.

\section{Acknowledgment}
The possibility of an error in Lemma 4.10 of \cite{DBLP:journals/siamcomp/BaswanaGS15} was pointed out by Sayan Bhattacharya and Divyarthi Mohan. 
The second author would like to thank both of them for discussions on the proof of the expectation bound and the definition of $B(v,i,k)$. 
In \cite{DBLP:journals/siamcomp/BaswanaGS15}, we used $2^i$ as the threshold for raising a vertex to level $i$. The possibility of increasing this threshold from $2^i$ to $b^i$ for any constant $b$ without any impact on the time complexity was observed by Shay Solomon \cite{DBLP:journals/corr/Solomon16}. We are thankful to him for this observation.
\bibliographystyle{plain}
\bibliography{sample}

\begin{thebibliography}{10}

\bibitem{AbboudW14}
Amir Abboud and Virginia~Vassilevska Williams.
\newblock Popular conjectures imply strong lower bounds for dynamic problems.
\newblock {\em CoRR}, abs/1402.0054, 2014.

\bibitem{AbrahamIKM07}
David~J. Abraham, Robert~W. Irving, Telikepalli Kavitha, and Kurt Mehlhorn.
\newblock Popular matchings.
\newblock {\em SIAM J. Comput.}, 37(4):1030--1045, 2007.

\bibitem{AnandBGS12}
Abhash Anand, Surender Baswana, Manoj Gupta, and Sandeep Sen.
\newblock Maintaining approximate maximum weighted matching in fully dynamic
  graphs.
\newblock In {\em FSTTCS}, pages 257--266, 2012.

\bibitem{AnandBGS-Arxiv12}
Abhash Anand, Surender Baswana, Manoj Gupta, and Sandeep Sen.
\newblock Maintaining approximate maximum weighted matching in fully dynamic
  graphs.
\newblock {\em CoRR}, abs/1207.3976, 2012.

\bibitem{DBLP:journals/siamcomp/BaswanaGS15}
Surender Baswana, Manoj Gupta, and Sandeep Sen.
\newblock Fully dynamic maximal matching in {O}(log n) update time.
\newblock {\em {SIAM} J. Comput. ({\em preliminary version appeared in FOCS
  2011})}, 44(1):88--113, 2015.

\bibitem{BaswanaKS12}
Surender Baswana, Sumeet Khurana, and Soumojit Sarkar.
\newblock Fully dynamic randomized algorithms for graph spanners.
\newblock {\em ACM Transactions on Algorithms}, 8(4):35, 2012.

\bibitem{DBLP:conf/icalp/BernsteinS15}
Aaron Bernstein and Cliff Stein.
\newblock Fully dynamic matching in bipartite graphs.
\newblock In {\em Automata, Languages, and Programming - 42nd International
  Colloquium, {ICALP} 2015, Kyoto, Japan, July 6-10, 2015, Proceedings, Part
  {I}}, pages 167--179, 2015.

\bibitem{DBLP:conf/soda/BernsteinS16}
Aaron Bernstein and Cliff Stein.
\newblock Faster fully dynamic matchings with small approximation ratios.
\newblock In {\em Proceedings of the Twenty-Seventh Annual {ACM-SIAM} Symposium
  on Discrete Algorithms, {SODA} 2016, Arlington, VA, USA, January 10-12,
  2016}, pages 692--711, 2016.

\bibitem{DBLP:conf/stoc/BhattacharyaHN16}
Sayan Bhattacharya, Monika Henzinger, and Danupon Nanongkai.
\newblock New deterministic approximation algorithms for fully dynamic
  matching.
\newblock In {\em Proceedings of the 48th Annual {ACM} {SIGACT} Symposium on
  Theory of Computing, {STOC} 2016, Cambridge, MA, USA, June 18-21, 2016},
  pages 398--411, 2016.

\bibitem{Edmonds65}
J.~Edmonds.
\newblock Paths, trees, and flowers.
\newblock {\em Canadian Journal of Mathematics}, 17:449–467, 1965.

\bibitem{EdmondsJohnson73}
J.~Edmonds and E.~L. Johnson.
\newblock Matching, euler tours, and the chinese postman.
\newblock {\em Mathematical Programming}, 5:88--124, 1973.

\bibitem{GabowTarjan91}
Harold~N. Gabow and Robert~Endre Tarjan.
\newblock Faster scaling algorithms for general graph-matching problems.
\newblock {\em J. ACM}, 38(4):815--853, 1991.

\bibitem{GaleShapley62}
D.~Gale and L.~S. Shapley.
\newblock College admissions and the stability of marriage.
\newblock {\em American Mathematical Monthly}, 69:9--14, 1962.

\bibitem{GuptaP13}
Manoj Gupta and Richard Peng.
\newblock Fully dynamic \$(1+)\$-approximate matchings.
\newblock In {\em 54th Annual IEEE Symposium on Foundations of Computer
  Science}, 2013.

\bibitem{DBLP:conf/stoc/HenzingerKNS15}
Monika Henzinger, Sebastian Krinninger, Danupon Nanongkai, and Thatchaphol
  Saranurak.
\newblock Unifying and strengthening hardness for dynamic problems via the
  online matrix-vector multiplication conjecture.
\newblock In {\em Proceedings of the Forty-Seventh Annual {ACM} on Symposium on
  Theory of Computing, {STOC} 2015, Portland, OR, USA, June 14-17, 2015}, pages
  21--30, 2015.

\bibitem{HolmLT01}
Jacob Holm, Kristian de~Lichtenberg, and Mikkel Thorup.
\newblock Poly-logarithmic deterministic fully-dynamic algorithms for
  connectivity, minimum spanning tree, 2-edge, and biconnectivity.
\newblock {\em J. ACM}, 48(4):723--760, 2001.

\bibitem{HopcroftKarp73}
John~E. Hopcroft and Richard~M. Karp.
\newblock An $n^{5/2}$ algorithm for maximum matchings in bipartite graphs.
\newblock {\em SIAM J. Comput.}, 2(4):225--231, 1973.

\bibitem{HuangKavitha12}
Chien-Chung Huang and Telikepalli Kavitha.
\newblock Efficient algorithms for maximum weight matchings in general graphs
  with small edge weights.
\newblock In {\em SODA}, pages 1400--1412, 2012.

\bibitem{KleinbergTardos05}
Jon Kleinberg and Eva Tardos.
\newblock {\em Algorithm Design}.
\newblock Addison Wesley, 2005.

\bibitem{Lawler76}
E.~Lawler.
\newblock {\em Combinatorial Optimization: Networks and Matroids}.
\newblock Holt, Rinehart \& Winston, Newyork, 1976.

\bibitem{LP86}
L.~Lovasz and M.D. Plummer.
\newblock {\em Matching Theory}.
\newblock AMS Chelsea Publishing, North-Holland, Amsterdam–New York, 1986.

\bibitem{MicaliVazirani80}
Silvio Micali and Vijay~V. Vazirani.
\newblock An {$O(\sqrt{(|V|)} |E|)$} algorithm for finding maximum matching in
  general graphs.
\newblock In {\em FOCS}, pages 17--27, 1980.

\bibitem{MuchaS04}
Marcin Mucha and Piotr Sankowski.
\newblock Maximum matchings via gaussian elimination.
\newblock In {\em FOCS}, pages 248--255, 2004.

\bibitem{NeimanS12}
Ofer Neiman and Shay Solomon.
\newblock Deterministic algorithms for fully dynamic maximal matching.
\newblock {\em CoRR}, abs/1207.1277, 2012.

\bibitem{OnakRubinfeld10}
Krzysztof Onak and Ronitt Rubinfeld.
\newblock Maintaining a large matching and a small vertex cover.
\newblock In {\em STOC}, pages 457--464, 2010.

\bibitem{cuckoo}
Rasmus Pagh and Flemming~Friche Rodler.
\newblock Cuckoo hashing.
\newblock {\em J. Algorithms}, 51(2):122--144, 2004.

\bibitem{RodittyZwick08}
Liam Roditty and Uri Zwick.
\newblock Improved dynamic reachability algorithms for directed graphs.
\newblock {\em SIAM J. Comput.}, 37(5):1455--1471, 2008.

\bibitem{RodittyZwick12}
Liam Roditty and Uri Zwick.
\newblock Dynamic approximate all-pairs shortest paths in undirected graphs.
\newblock {\em SIAM J. Comput.}, 41(3):670--683, 2012.

\bibitem{DBLP:journals/corr/Solomon16}
Shay Solomon.
\newblock Fully dynamic maximal matching in constant update time.
\newblock {\em CoRR}, abs/1604.08491, 2016.

\bibitem{Thorup07}
Mikkel Thorup.
\newblock Fully-dynamic min-cut.
\newblock {\em Combinatorica}, 27(1):91--127, 2007.

\end{thebibliography}

\section{Appendix}
\subsubsection*{Proof of Lemma \ref{lemma:conditional-probability}}
\begin{proof}
\begin{eqnarray*}
\PP[A\cap C] & = & \PP[A\cap (\cup_i B_i)] \\
                    & = & \sum_i \PP[A\cap B_i]  ~~~~~~~~\mbox{since $B_i$'s are mutually exclusive} \\
                    & = & \sum_i \PP[A~|~B_i]\cdot \PP[B_i] ~~~~~~~~\mbox{using the definition of conditional probability}\\
                    & = & \rho \cdot \sum_i \PP[B_i] \\
                    & = & \rho \cdot \PP[\cup_i B_i]  = \rho \cdot \PP[C]~~~~~~~~ \mbox{since $B_i$'s are mutually exclusive}
\end{eqnarray*}
Hence $\PP[A~|~C] ~=~\PP[A\cap C] / \PP[C] ~=~ \rho$.
\end{proof}

\subsubsection*{Proof of Lemma \ref{lemma:independence-is-symmetric-lemma}}

\begin{proof}
Since $A$ is independent of $X$, so for each $x\in X$,
\begin{equation} 
         \PP[A \cap X=x] = \PP[A] \cdot \PP[X=x] 
\label{eq:product-of-independent-events}
\end{equation}
Hence 
\begin{eqnarray*}
\PP[X=x | A] & = & \frac{\PP[A \cap X=x]}{\PP[A]}\\
                   & = & \frac{\PP[A] \cdot \PP[X=x]}{\PP[A]}  ~~~~\mbox{using Equation \ref{eq:product-of-independent-events}} \\
                   & = & \PP[X=x]   
\end{eqnarray*}
\end{proof}

\subsubsection*{Proof of Lemma \ref{lemma:asymmetric-random-walk}}
\noindent

The asymmetric random walk problem can be seen as a special case of the famous Gambler's ruin problem described as follows.

\noindent
{\bf Gambler's ruin problem.}\\
There are two players who play a game that goes in rounds. Initially Player $1$ has a capital of $c$
units and Player $2$ has a capital of $c'$ units. Player $1$ wins a round with probability $p$ and loses with probability $q=1-p$ independent of the previous rounds. The winner of a round takes away one unit of the capital from the opponent. 
The game ends when the capital of one of the players becomes 0. 

The following lemma is well-known in many text books on probability theory. A concise and self contained proof is available at the link : {\footnotesize{examplehttp://faculty.washington.edu/fm1/394/Materials/Gambler.pdf}}.

\begin{lemma}
In the Gambler's ruin problem with $p>q$, the probability that Player $1$ gets ruined is 
\[ 
 \frac{1-(p/q)^{c'}}{1-(p/q)^{c+c'}}
\]
\label{lemma:gambler's-ruin}
\end{lemma}

Let us now put an additional restriction in the problem: the total number of rounds allowed in the game is $L$ for a given number $L<c'$.
Notice that with this restriction Player 2 will never be ruined. As a result the game will be over when Player 1 gets ruined or
when $L$ rounds are over. The probability of Player 1 getting ruined in this restricted Gambler's problem is strictly
less than the probability of Player 1 getting ruined in the original Gambler's problem described above. This is because there is a 
non-zero probability that Player 1 may be ruined after performing more than $L$ steps, and the restricted Gambler's problem rules out this
possibility. Hence, using Lemma \ref{lemma:gambler's-ruin}, it follows that for the restricted version of the Gambler's ruin problem with $p>q$, the probability that Player $1$ gets ruined is less than
\[ 
 \frac{1-(p/q)^{c'}}{1-(p/q)^{c+c'}} < \left(\frac{q}{p}\right)^c
\]
The restricted version of the Gambler's ruin problem can be formulated as an asymmetric random walk problem:
The walk starts at location $c$ units to the right of the origin. In each step, the particle moves one unit to the right with probability $p$
or one unit to the left with probability $q=1-p$ indepependent of the past moves. The walk terminates upon reaching either the origin or when it has performed $L$ step. This completes the proof of Lemma \ref{lemma:asymmetric-random-walk}.

\end{document}